\documentclass{emulateapj}

\newcommand{\etal}{{\it et\thinspace al.}\ }

\newcommand{\simlt}{\ {\raise-.5ex\hbox{$\buildrel<\over\sim$}}\ }

\submitted{Submitted 6 November 2004; Accepted 3 February 2005 -- To appear in 20 May 2005 ApJ}

\lefthead{Salzer \etal}
\righthead{Optical and IR L-Z Relations}

\begin{document}

\title{Metal Abundances of KISS Galaxies.  IV.\\ 
Galaxian Luminosity-Metallicity Relations in the Optical and Near-IR}

\author{John J. Salzer}
\affil{Astronomy Department, Wesleyan University, Middletown, CT 06459; slaz@astro.wesleyan.edu}

\author{Janice C. Lee}
\affil{Steward Observatory, University of Arizona, 933 North Cherry Avenue, Tucson, AZ 85721; jlee@as.arizona.edu}

\author{Jason Melbourne}
\affil{UCO/Lick Observatory, UC Santa Cruz, Santa Cruz, CA 95064; jmel@ucolick.org}

\author{Joannah L. Hinz and Almudena Alonso-Herrero\altaffilmark{1}}
\affil{Steward Observatory, University of Arizona, 933 North Cherry Avenue, Tucson, AZ 85721; jhinz@as.arizona.edu; aalonso@as.arizona.edu}

\and

\author{Anna Jangren}
\affil{Astronomy Department, Wesleyan University, Middletown, CT 06459; anna@astro.wesleyan.edu}

\altaffiltext{1}{Present address: Departamento de Astrof\'{\i}sica Molecular e 
Infrarroja, IEM, Consejo Superior de Investigaciones Cient\'{\i}ficas, Serrano 113b, 
28006 Madrid, Spain.}

\clearpage

\begin{abstract}
We explore the galaxian luminosity-metallicity (L-Z) relationship in both the optical 
and the near-IR using a combination of optical photometric and spectroscopic observations 
from the KPNO International Spectroscopic Survey (KISS) and near-infrared photometry 
from the Two-micron All Sky Survey (2MASS).  We supplement 
the 2MASS data with our own NIR photometry for a small number of lower-luminosity 
ELGs that are under-represented in the 2MASS database.  Our B-band L-Z relationship 
includes 765 star-forming KISS galaxies with coarse abundance estimates from our 
follow-up spectra, while the correlation with KISS and 2MASS yields a total of 420 
galaxies in our J-band L-Z relationship.  We explore the effect that changing the 
correlation between the strong-line abundance diagnostic R$_{23}$ and metallicity 
has on the derived L-Z relation.  We find that the slope of the L-Z relationship 
decreases as the wavelength of the luminosity bandpass increases.  We interpret this 
as being, at least in part, an effect of internal absorption in the host galaxy.  
Furthermore, the dispersion in the L-Z relation decreases for the NIR
bands, suggesting that variations in internal absorption contribute significantly
to the observed scatter.  We propose that our NIR L-Z relations are more fundamental
than the B-band relation, since they are largely free of absorption effects and
the NIR luminosities are more directly related to the stellar mass of the galaxy
than are the optical luminosities.
\end{abstract}


%************************************************************************

\section{Introduction}

Galaxies build up their content of heavy elements slowly over their lifetimes
through well understood stellar-evolution processes.   It has been known for
at least twenty-five years that galaxies can exhibit different metallicity levels, 
with more massive galaxies having systematically higher abundances than 
dwarf galaxies (Lequeux \etal 1979).   Understanding
why these differences in metallicity exist is an important aspect of the study of
galaxy evolution.   The accurate assessment of the relationship between
metallicity and mass provides an important constraint on the models of
galaxy formation and evolution that attempt to account for the chemical
evolution of the system (e.g., Prantzos \& Boissier 2000; Boissier \etal 2003;  
Chiappini, Romano \& Matteucci 2003;  Rieschick \& Hensler 2003; Qian \& 
Wasserburg 2004). 

While it would be preferable to study the variation of metallicity with the mass 
of the galaxy, it is far more common for investigators to use luminosity as a 
surrogate for mass (e.g., Garnett \& Shields 1987; Skillman, Kennicutt, \& Hodge 
1989; Zaritsky, Kennicutt, \& Huchra 1994; Richer \& McCall 1995; Garnett \etal 
1997; Hidalgo-G\'amez \& Olofsson 1998; Kobulnicky \& 
Zaritsky 1999; Pilyugin \& Ferrini 2000; Melbourne \& Salzer 2002;  Lee \etal 2004; 
Lamareille \etal 2004).  In nearly all of
these previous studies, the authors find a strong correlation between luminosity
and heavy element content.  The latter is usually specified in terms of the
relative abundance of oxygen to hydrogen for late-type and star-forming
galaxies (where the metallicity is measured via nebular emission lines), and
in terms of Fe/H for early-type galaxies (where the metallicity is assessed in
terms of absorption-line indices).  A few recent studies have considered the
relationships between circular rotation speed and metallicity (Garnett 2002)
and spectroscopic/photometric stellar mass estimates and metallicity (Tremonti 
\etal 2004, P\'erez-Gonz\'alez \etal 2003b), in order to try to tie the observed 
variations in metal abundance to a  more fundamental parameter (e.g., stellar 
or total mass).

There are a number of possible explanations for the observed variation of metal content 
with galaxian mass.  For example, dwarf galaxies could, on average, be less 
efficient at converting their gas into stars, both now and in the past, and hence 
could simply have produced lower amounts of heavy elements in their lifetimes.
Alternatively, it is possible that differences in the initial mass functions of 
galaxies as a function of
either their metallicity or total mass could explain the observed trends.
Both of these possibilities are brought into question by the observations of
actively star-forming dwarf galaxies (a.k.a., blue compact dwarfs), that are seen
to contain many high-mass stars capable of producing high levels
of metal enrichment.  For any reasonable assumed duty cycle of the star
formation episodes in these galaxies, it would appear that they would readily
produce enough heavy elements to give themselves relative abundances
comparable to more massive galaxies.  A key wild-card in the star-formation and
metal enrichment in dwarfs is the observation that they have high gas-mass
fractions relative to larger galaxies.  Most of their  ``excess" gas lies at large
radii, outside of the optical extent of the galaxy.  If a mechanism existed to
mix some of this gas into the central portion of the galaxy (e.g., tidal forces),
it would serve to dilute the metallicity of the gas in the observable portion of
the galaxy, and also to fuel further star formation.

A more likely explanation for the observed mass-metallicity and luminosity-metallicity
(L-Z) relations is preferential loss of metal-enriched supernova ejecta.  Low mass
galaxies, with lower gravitational binding energies, are more apt to lose the
metals produced in SNe explosions than are more massive galaxies.  One can also
explain the observed metallicity gradients in spiral disks via this mechanism.
Simulations by Mac Low \& Ferrara (1999) have shown that
above some critical mass (roughly corresponding to a galaxy with M$_B$ = $-$12),
galaxies will lose a substantial amount of the metals they produce via a hot
galaxian wind, while leaving most of the colder, ambient gas intact for the next
generation of star formation.   The extent of the metal loss via this supernova-driven
wind varies with the mass of the galaxy, such that dwarfs lose much of their produced
metals, while large spirals lose little if any.  While the precise mechanism that
causes the L-Z relation remains an open question, we will assume in the remainder 
of this paper, for the sake of discussion, that the hot galaxian wind model is the primary 
physical mechanism that produces the observed correlation between metallicity and
luminosity/mass.

Recent work on the L-Z relation has begun to focus on trying to measure the
change in the relation as a function of look-back time by studying galaxies at
moderate redshifts (Kobulnicky \& Zaritsky 1999; Contini \etal 2002;
Kobulnicky \etal 2003; Lilly, Carollo \& Stockton 2003; Maier, Meisenheimer \&
Hippelein 2004; Liang \etal 2004).  Studying the evolution of the L-Z relation can be 
an alternate method for trying to quantify the evolution of the star-formation rate
density.  Since the evolution of metal abundance, particularly in high-mass
galaxies where the effect of galaxian winds is less important, must correlate
with the star-formation history of massive stars, the evaluation of the change
in the L-Z relation with look-back time provides a powerful tool for measuring
the star-formation history of the universe back to z = 1 (and possibly further).

Melbourne \& Salzer (2002; Paper I) have previously used the star-forming galaxies discovered 
in the KPNO International Spectroscopic Survey (KISS; Salzer \etal 2000) to explore the
nature of the L-Z relation for a large (N = 519) sample of emission-line galaxies (ELGs).  
In the current paper, we extend that study by including a larger sample of galaxies
for which coarse abundance estimates are available, and by investigating the L-Z
relation in a range of bandpasses from B to K.  By considering the L-Z relation in the near-IR,
we are able to substantially reduce (eliminate?) the effects of internal absorption on
the derived result.  In addition, we have revised our coarse abundance calibration,
taking into account recently obtained spectral data.  An added advantage of the
current study is that we use a large sample of galaxies with {\it homogeneous} data
(both spectroscopic and photometric).  Some previous studies have suffered from
using data taken from multiple sources, where the abundances were not always
derived in a consistent fashion.

The paper is laid out as follows.
Section 2 describes the various optical and NIR datasets used for this study, including
the presentation of new H and K band imaging photometry obtained for this project.
Section 3 details our method of deriving coarse metallicity estimates for the KISS
ELGs, and includes a redetermination of the metallicity - line ratio relations first
presented in Paper I.  In Section 4 we present our new L-Z
relations for five photometric bands (BVJHK), while \S 5 presents a discussion of
our findings.  These include a comparison with previous work, including recent
determinations of the L-Z relation at higher redshifts, as well as a discussion of
the scatter in our relation.  The results of this paper are summarized in \S 6.
Throughout this paper, a value for the Hubble constant of 75 km s$^{-1}$ Mpc$^{-1}$
is assumed.

%************************************************************************

\section{Observational Data}

To explore the questions raised above, we utilize the large sample of star-forming 
galaxies cataloged by the KPNO International Spectroscopic Survey (KISS; Salzer \etal 
2000).   To date, 
KISS has generated three lists of emission-line selected galaxies.  The larger portion
of the survey is made up of galaxies selected via the presence of the H$\alpha$ line
in their objective-prism spectra.  These are cataloged in Salzer \etal (2001; hereafter
KR1) and Gronwall \etal (2004a; hereafter KR2), and include 2158 emission-line
galaxy (ELG) candidates.  The third list is made up of objects selected due to the
presence of [\ion{O}{3}]$\lambda$5007 (Salzer \etal 2002; hereafter KB1), and has a 
total of 223 objects cataloged.  The [\ion{O}{3}]-selected ELGs in KB1 tend to be
lower in luminosity, on average, than the H$\alpha$-selected galaxies in KR1 and KR2.
However, there is substantial overlap in the sky coverage of the KR1 and KB1 surveys,
and essentially all of the KB1 objects in the overlap regions are also detected in
KR1.  This suggests that there is no significant bias introduced by combining the
H$\alpha$-selected and [\ion{O}{3}]-selected ELGs.  The total number of unique ELGs 
tabulated in the three lists is 2266.

The current study makes use of a large amount of observational data, obtained from
a number of different sources.  We describe below the make-up of our dataset.

\subsection{KISS Optical Photometry and Spectroscopy}

The basic data in the KISS catalogs include accurate astrometry (typical 
uncertainties of 0.25 - 0.30 arcsec in RA and Dec) and calibrated B \& V photometry
for each object in the catalog.  The latter have typical uncertainties of 0.06 mag for 
B = 16 galaxies, increasing to 0.09 mag at B = 19 (Salzer \etal 2000).  In addition, the 
objective-prism data provide estimates of the redshift of each ELG, plus a measurement 
of its line flux and equivalent width.  These spectral data are too coarse to allow for the 
determination of activity types (e.g., star-forming vs. AGN), so follow-up spectra 
are needed to maximize the scientific usefulness of the survey.  Members of the KISS 
group have been engaged in a campaign of ``quick-look" spectroscopy of the ELG 
candidates for the past several years.  To date, observations have been obtained for 
1351 KISS ELGs, including all 223 of the [\ion{O}{3}]-selected KB1 list, 935 of 1128 
(83\%) ELGS from the H$\alpha$-selected KR1 catalog, and 307 of 1029 (30\%) of the 
candidates from KR2.  For the analysis carried out in this paper, we use only
spectra from star-forming galaxies that have spectral data of high enough quality
to insure accurate emission-line ratios (i.e., galaxies with spectral quality codes
of 1 or 2 as tabulated in the spectral data papers listed below).  After removing
all of the AGNs (Seyfert 1s and 2s, LINERs, and QSOs), those objects with lower 
spectral quality, and the roughly 9\% of KISS candidates that turn out to not be
actual ELGs, we are left with a final sample for the current paper of 766 star-forming
ELGs.  

The follow-up spectra have been obtained with a variety of telescopes and 
spectrographs.   The details of the observational set-ups and measured spectral
parameters are given in a series of existing or planned future papers, each of which
presents data from a specific telescope.  These include the Hobby-Eberly 9.2-m telescope 
(Gronwall \etal 2004b), the Lick 3.0-m (Melbourne \etal 2004; Paper II), the MDM 2.4-m 
(Wegner \etal 2003), the WIYN 3.5-m with the Hydra multi-fiber positioner, the ARC 3.5-m, 
and the KPNO 2.1-m telescope.   With the exception of the WIYN 3.5-m spectra, all data 
were obtained using long slits with widths of 1.5 -- 2.0 arcsecs.  The WIYN spectra 
were taken through 2 arcsec diameter fibers. 
The quick-look spectra exhibit a range of S/N ratios.  Some spectra
cover the full optical spectral range (from below [\ion{O}{2}]$\lambda\lambda$3726,3729 
to above [\ion{S}{2}]$\lambda\lambda$6717,6731), although the majority of our data do
not go below 4000 \AA, usually due to limitations of the spectrographs that were available.
At a minimum, all of our spectra include the H$\beta$, H$\alpha$, 
[\ion{O}{3}]$\lambda\lambda$4959,5007, and [\ion{N}{2}]$\lambda\lambda$6548,6583 lines.
In addition, the follow-up spectra provide us with an estimate of the Balmer decrement reddening
parameter c$_{H\beta}$, plus an accurate redshift.  The latter, along with our broad-band
photometry from the original survey data, allow us to compute absolute magnitudes for
each ELG.  The emission-line fluxes are used to estimate the oxygen abundance, using
the method described in \S 3.

\subsection{2MASS NIR Photometry}

In order to explore the near-IR L-Z relationship, we take advantage of the 
Two-micron All-Sky Survey\footnote{This publication makes use of data products from
the Two-micron All-Sky Survey, which is a joint project of the University of Massachusetts
and the Infrared Processing and Analysis Center/California Institute of Technology, funded
by the National Aeronautics and Space Administration and the National Science Foundation.} 
(2MASS).  We have correlated the positions of
all 2266 KISS ELGs in the first three survey lists (KR1, KR2, KB1) with both the 2MASS
point-source and extended-source catalogs.  We have found 2MASS detections for 1479 of the
KISS ELGs (65.3\%).  In most cases, the KISS galaxies were detected in all three 2MASS
bands (J, H, and K), although for some sources near the 2MASS flux limit the KISS ELGs
were detected in only one (usually J) or two (usually J \& H) filters.  The median formal
magnitude uncertainties for the 2MASS-detected KISS galaxies are 0.10 mag for J, 0.13 mag
for H, and 0.15 mag for K.  The full details of our correlation procedure, as well as a 
description of the NIR properties of the KISS ELGs is given in A. Jangren \etal (2005b, in 
preparation).  In the current paper we simply utilize the cataloged 2MASS 
JHK photometry to compute NIR absolute magnitudes for all of the detected ELGs.

\subsection{Steward NIR Observations}

As discussed in A. Jangren \etal (2005b, in preparation), the 2MASS survey was quite 
efficient at detecting the more luminous star-forming galaxies and AGNs in the KISS 
catalogs.  However, the lower luminosity ELGs are systematically under-represented in 
2MASS.  This comes about for two reasons.  First, since the detection limit of the 2MASS 
project is not as deep as
that for KISS, a substantial portion of the KISS dwarf population is not within the 
sensitivity limits of 2MASS.  Second, the colors of the low-luminosity star-forming 
ELGs are much bluer than are the more luminous KISS galaxies.  Hence, at a given 
optical magnitude, the more luminous (redder) ELGs are more likely to by detected 
in the NIR than the bluer dwarf galaxies.  The net effect is that the majority of 
the lower luminosity KISS ELGs (M$_B$ $>$ -18.0) are not detected by 2MASS.  

To supplement the sparse 2MASS photometry available for the low-luminosity galaxies
of the KISS sample, we have carried out near-IR observations for some of our lowest 
luminosity galaxies.  Our original intent was to obtain both $K_s$ and $H$ 
band imaging for all 52 galaxies with $M_B > -17.0$ in the KR1 catalog.  
Due to unusual difficulties with weather, however, only a fraction of this 
sample was observed.  For the purposes of constructing the luminosity-metallicity 
relationship using coarse abundances, the inclusion of this limited set of near-IR 
photometry is still sufficient to improve our statistics in the low-luminosity 
portion of the distribution.   Thus, we have simply abridged the original project and 
present the available data here. 

H-band images were obtained with a NICMOS3 256 $\times$ 256 camera at the 
Steward Observatory 2.3m Bok Telescope on Kitt Peak in June 2002.  This 
MOSFET array has a pixel scale of 0\farcs60 pixel$^{-1}$ and a 2\farcm56 
square field of view when used at the f/45 focus.

$K_s$ images were obtained with the PISCES camera (McCarthy et al. 2001) at 
the 6.5m MMT\footnote{The MMT Observatory is a joint facility of the University of Arizona and 
the Smithsonian Institution.} on Mount Hopkins in April 2003.   PISCES uses a 1024 $\times$ 1024 
pixel HAWAII (HgCdTe) detector (Kozlowski et al. 1994; Hodapp et al. 1996) and operates at the 
f/9 focus at the MMT, with a resultant pixel scale of 0\farcs185 pixel$^{-1}$ and circular field 
of view with a 3\farcm16 diameter.

At both observatories we followed the same standard near-IR observing sequence.  
A series of on-source short exposures were taken for each target, with dithering 
offsets applied after every frame and the integration time ($\sim$30s) of each 
frame determined by the level of the sky background count rate.  Since the 
targets had diameters on the order of 10\arcsec\ or smaller and the surrounding 
fields were relatively sparse, it was not necessary to move the target completely 
off the detector to obtain separate blank sky images for flat-fielding and sky-subtraction 
purposes.  Total integration times varied between 10 and 20 minutes for each target.

Standard near-IR reductions were performed using the Image Reduction and Analysis 
Facility\footnote{IRAF is distributed by the National Optical Astronomy 
Observatories, which are operated by AURA, Inc.\ under cooperative agreement
with the National Science Foundation.} (IRAF).  Average dark frames were 
subtracted, bad pixel masks were applied, and the images were flat-fielded 
using a normalized median combination of the 5 to 10 temporally closest 
frames.  For the PISCES data, corrections for cross-talk between quadrants 
and geometric distortion were also applied.  Finally, the individual short 
exposure frames for each target were aligned and combined.  Flux calibration 
was performed using standard stars from the UKIRT list for data from the Bok 
Telescope and from the Persson et al. (1998) list for data from the MMT, and 
the canonical atmospheric extinction coefficients of 0.04 mag airmass$^{-1}$ 
at $H$ and 0.08 mag airmass$^{-1}$ at $K_s$ were used to calculate zero-points 
for each night.  The uncertainty in the airmass correction was minimized by 
ensuring that all standard and galaxy observations were made at low airmass 
(1.2 airmasses on average) and over the narrowest range of airmass that was 
practical ($\sim$0.25).

Photometry was performed using the IRAF APPHOT routine and total magnitudes 
were measured by using circular apertures large enough to encompass all the 
light from the target.  Our results are listed in Table~\ref{tab:nirphot}.  The quoted errors 
were computed following P\'erez-Gonz\'alez et al. (2003a).  Average total errors are 0.07 
mag for the MMT K$_s$ photometry and 0.3 mag for the Bok H-band photometry.  
The larger errors in the Bok data are primarily due to a greater uncertainty in 
the sky determination due to the high and rapidly fluctuating background count 
rate at the time of the observations.  For the MMT PISCES data, we have also 
measured field stars in our science images for which reliable photometry is 
also available from the 2MASS PCS.  There were typically one to two stars per 
field that could be used for this type of comparison.  The differences between 
our measured values and those from 2MASS were consistently within 1.5$\sigma$ 
of the quoted 2MASS errors, validating the photometric calibrations that were 
applied.

%************************************************************************

\section{Metallicities of the KISS Galaxies}

The metallicity (defined here as 12 + log(O/H)) of nebular star-forming
regions can be calculated from ratios of the [\ion{O}{3}] and [\ion{O}{2}] 
spectral lines relative to a Balmer line of hydrogen (Osterbrock 1989; de 
Robertis 1987).  This requires high
signal-to-noise ratio spectra and the detection of the oxygen auroral line
[\ion{O}{3}]$\lambda$4363 to estimate electron temperature in the nebula.  We 
term this abundance measurement method the $T_e$ method.  The strength of 
the auroral line is usually 50-200 times weaker than [\ion{O}{3}]$\lambda$5007, 
making it difficult to observe.  To make matters worse, if the metallicity 
of the star-forming region is larger than 12 + log(O/H) $\sim8$, the [\ion{O}{3}] 
line strengths and the [\ion{O}{3}]$\lambda$4363/[\ion{O}{3}]$\lambda$5007  
ratio decline with increasing oxygen abundance, making the detection 
of the auroral line impossible for high metallicity targets at modest redshifts.  
In order to estimate the metallicity for cases where [\ion{O}{3}]$\lambda$4363 is 
not observed, empirical relations between ratios of the strong oxygen lines 
([\ion{O}{3}]$\lambda\lambda$5007,4959 and [\ion{O}{2}]$\lambda\lambda$3726,3729) 
and metallicity are calibrated using HII regions in the Milky Way and local galaxies.  
These so-called $R_{23}$ methods (Edmunds and Pagel 1984; Zaritsky at al. 1994; 
Kobulnicky et al. 1999; Pilyugin 2000, 2001) have typical metallicity uncertainties 
of $\sim$0.1--0.2 dex, as compared to uncertainties of $\sim$0.01--0.04 dex that
are typically obtained via the $T_e$ method.   The recent study by Kewley \& Dopita 
(2002) has refined the strategies for deriving metal abundances making use of the 
strongest nebular lines present in the spectrum of star-forming galaxies.
 
The follow-up spectra of a typical KISS emission-line galaxy generally contain 
detections of the H$\alpha$, H$\beta$, [\ion{O}{3}]$\lambda$5007, and 
[\ion{N}{2}]$\lambda6583$ lines.  However, due to limited spectral coverage and
to the sometimes low signal-to-noise ratio (S/N) of the quick-look spectra, they 
often do not contain the additional lines necessary for accurate metal abundances, 
such as [\ion{O}{3}]$\lambda$4363 ($T_e$ abundances) or [\ion{O}{2}]$\lambda\lambda$3726,3729 
($T_e$ and $R_{23}$ abundances).   In particular, the absence of the [\ion{O}{2}] doublet
precludes our using any of the strong-line ratios proposed by
Kewley \& Dopita (2002).  Rather than abandon these objects, we define an 
empirical calibration between metallicity and the strong lines available in the bulk 
of our spectra.  These calibrations result from metallicity measurements of a subset
of the KISS galaxies for which we can either calculate a $T_e$ abundance or an $R_{23}$ 
abundance.  The process was first carried out in Paper I. In the 
current paper we update the calibrations with additional $T_e$ and $R_{23}$ abundance
measurements, and we apply the calibrations to a larger set of follow-up spectra in 
order to determine global luminosity-metallicity relations for KISS galaxies.  A 
brief description of the data sets used and their application to the strong-line 
metallicity calibration is given below.  For a more detailed discussion of the process 
please refer to Paper I.  

\subsection{$T_e$ Metal Abundances}

The KISS team has obtained high S/N spectra of 25 low metal abundance galaxies
using the Lick 3.0-m telescope and the 6.5-m MMT.  Each 
spectrum includes a detection of [\ion{O}{3}]$\lambda$4363, which we use to measure
the electron temperature of the nebula.  An estimate of the electron density comes 
from the [SII] line ratio.  Metal abundances are calculated using the IRAF NEBULAR 
package (de Robertis et al. 1997; Shaw \& Dufour 1995).  The abundance results are 
presented in a series of papers (Melbourne \etal 2004 (Paper II); Lee \etal 2004 (Paper III)), and are 
summarized in Table \ref{tab:temet}.  Columns 1 and 2 give the KISSR and/or KISSB numbers 
of the galaxies.  Column 3 shows the logarithm  of the ([\ion{N}{2}]$\lambda$6583/H$\alpha$)
line ratios, while columns 4 and 5 are log([\ion{O}{2}]$\lambda\lambda$3726,3729/H$\beta$) 
and log([\ion{O}{3}]$\lambda$5007/H$\beta$) respectively.  Column 6 gives the resulting 
$T_e$ metallicity.  Please refer to Papers II and III for full details.

\subsection{$R_{23}$ Metal Abundances}

In addition to the high signal-to-noise ratio spectra discussed above, the KISS 
team has gathered spectra of 185 galaxies that contain both the strong [\ion{O}{3}] 
and [\ion{O}{2}] lines allowing an $R_{23}$-style metallicity measurement for each.  
The relation between \\ $R_{23} = \frac{f([O III]\lambda\lambda4959,5007) + 
f([O II]\lambda\lambda3727,29)}{f(H\beta)}$ and metallicity has been calibrated with
a combination of $T_e$ observations of starburst galaxies and star-forming regions in 
nearby galaxies for the lower abundances, and photoionization models for the higher
metallicities (e.g., Stasinska 1990; McGaugh 1991; Oey \& Kennicutt 1993; Kewley \& 
Dopita 2002).  The relationship is double valued.  On the 
lower-metallicity branch (12 + log(O/H) $<7.9$), metallicity increases as $R_{23}$ 
increases.  $R_{23}$ decreases with increasing metallicity for metal abundances 
greater than $\sim8.1$, the upper-metallicity branch.  Objects in between are called 
turn-around region galaxies. 

After calculating $R_{23}$ for each object in
our sample we identify which branch of the metallicity relation the galaxy
lies on.  As outlined in Paper I, objects with 
log([\ion{N}{2}]$\lambda$6583/H$\alpha$) $<$ -1.3 are assigned to the low-metallicity
branch, while objects with log([\ion{N}{2}]$\lambda$6583/H$\alpha$) $>$ -1.0 are 
assigned to the high metallicity branch.  Those objects with 
nitrogen line ratios between these constraints are assigned to the 
turn-around region of the $R_{23}$ relation and no metallicity is calculated.
As an additional constraint, we find that objects with low signal-to-noise 
ratio in the H$\beta$ line do not produce reliable metal abundances from 
$R_{23}$ methods.  The uncertainty in H$\beta$ propagates through the 
metallicity determination by means of the reddening coefficient $c_{H\beta}$.  
We limit our metallicity calculations to those objects with equivalent
width of H$\beta > 8$ \AA.  Objects below this cutoff tend to have 
unphysical [\ion{O}{3}]/[\ion{O}{2}] line ratios (Paper I).  

\subsubsection{Lower-Metallicity Branch Calibration}

For objects on the lower-metallicity branch, we adopt the Pilyugin (2000) calibration
of $R_{23}$ called the $p3$ method:
\begin{equation}
12 + log(O/H) = 6.35 +1.45  X_3^*,
\end{equation}
where $X_3^* = -1.20 log(f([O III]\lambda\lambda4959,5007)/f($H$\beta)) + 
2.20 log(R_{23})$.  This equation reproduces $T_e$ abundance results to 
within 0.1 dex.  We calculate $p3$ metallicities for 17 galaxies with
metal abundances that range from 7.6 to 8.2.   The abundance results are
given in Table \ref{tab:p3met}.  Columns 1 and 2 give the KISSR and 
KISSB numbers of the galaxies.  Column 3 is 
log([\ion{N}{2}]$\lambda$6583/H$\alpha$), while columns 4 and 5 are
log([\ion{O}{2}]$\lambda\lambda$3726,3729/H$\beta$) and 
log([\ion{O}{3}]$\lambda$5007/H$\beta$) respectively.  Column 6 gives the
resulting $p3$ metallicity.  We assign a standard uncertainty to the
metallicity measurements of 0.1 dex (Paper I).

\subsubsection{Upper-Metallicity Branch Calibrations}

For objects on the upper-metallicity branch, we originally adopted  the Edmunds \& 
Pagel (1984; hereafter EP) fit, as quoted in Pilyugin (2000):
\begin{equation}
  12 + log(O/H) = 9.57  - 1.38 log(R_{23}).
\end{equation}
This fit differs only marginally from the one presented by Pilyugin (being $\sim$0.08
dex higher).  Table \ref{tab:R23met} reports the metallicity results for the $R_{23}$ 
method.  47 galaxies are measured with metal abundances ranging from 8.2 to 9.5.  The 
first six columns are the same as in Table \ref{tab:p3met}, except that the last
column reports the metal abundances derived from the $R_{23}$ equation above.  
The $R_{23}$ calibration has a larger inherent scatter with metallicity than does the 
$p3$ calibration; we estimate an uncertainty of 0.15 dex for these measurements 
(Paper I).

\begin{figure*}[htp]
\epsfxsize=5.0in
\epsscale{0.8}
\plotone{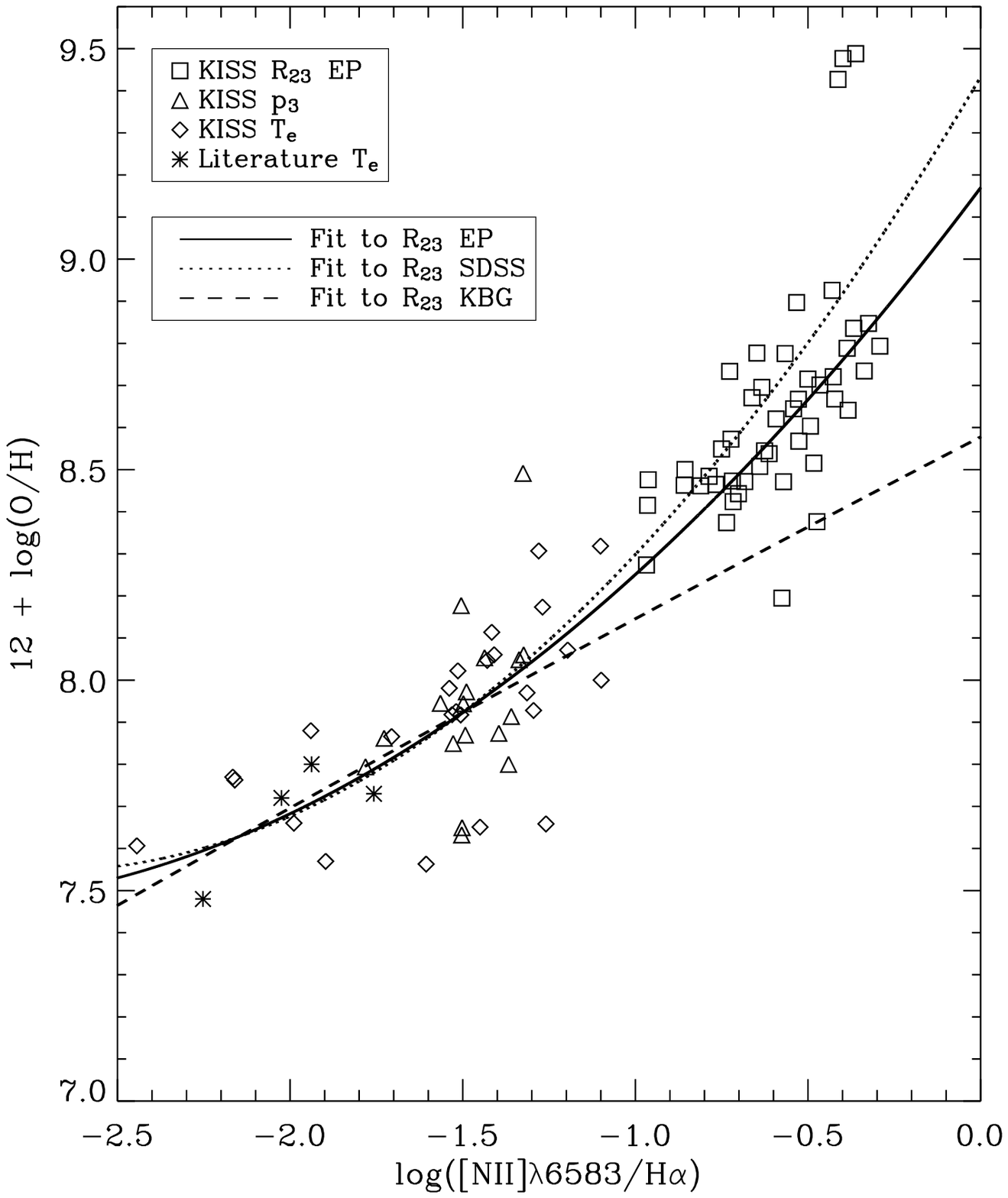}
\vskip -0.01in
\figcaption{This figure relates the [\ion{N}{2}]$\lambda$6583/H$\alpha$ line ratio to 
metallicity. Abundances calculated using the EP $R_{23}$ fit are shown as squares, 
while $p_3$ abundances are shown as triangles.  Abundances calculated with the 
T$_3$ method are shown as diamonds.  T$_e$ abundance data taken from Izotov \etal 
(1997) are shown as asterisks. We fit a quadratic function to the data (solid line).  In
addition, we plot the quadratic fits to the abundance data computed using the SDSS 
and KBG $R_{23}$ calibrations (dotted line and dashed line, respectively).  See 
text for details.
\label{fig:n2met}}
\end{figure*}

Our adopted upper-branch R$_{23}$ relationship defines the abundance scale for the luminous
galaxies used in the analysis presented in this paper.   The use of a different R$_{23}$ 
relationship could potentially have a dramatic impact on the slope of the derived L-Z 
relation.  It should  be stressed that there are a number of different calibrations of 
the R$_{23}$ relation in the literature (e.g., EP; McCall, Rybski \& Shields 1985; 
McGaugh 1991; Zaritsky \etal 1994; Kobulnicky \etal 1999; Pilyugin 2000, 2001), 
some of which differ significantly from each other.   We have adopted the EP
version, which falls roughly mid-range of the other versions of the relation. 
Paper I considered the effect that using a different R$_{23}$ relation
would have on the slope of the L-Z relation, and concluded that it could have only
a modest effect (probably in the range of 5-10\%).  However, we felt that it was important
to revisit this question in somewhat more detail by considering a few additional R$_{23}$
relations.

Recent measurements of T$_e$ abundances for objects on the upper branch of the 
R$_{23}$ relationship by Kennicutt, Bresolin \& Garnett (2003; hereafter KBG) lie well 
below the R$_{23}$ calibrations suggested in previous studies, including the one by
EP adopted above (see Figure 12 of KBG).  The implication of
this result is that abundances derived using one of these older R$_{23}$ relationships
could be substantially higher than their actual values.  We used the data presented in
KBG to derive an alternate fit for the upper branch of the R$_{23}$ relationship:
\begin{equation}
  12 + log(O/H) = 9.31  - 1.38 log(R_{23}).
\end{equation}
Note that the slope is identical to the Edmunds \& Pagel value, but the intercept is
0.26 dex lower.  Since a lower abundance calibration for the R$_{23}$ relationship
could potentially lead to a substantially shallower L-Z slope, we decided to recompute
our coarse abundance calibration using this new relation.  The R$_{23}$ abundances
derived for our sample using the KBG relation are given in column 7 of Table  
\ref{tab:R23met}.  

Finally, we have used the analytical fit to the R$_{23}$ -- metallicity  relation 
obtained by Tremonti \etal (2004) for SDSS data (their Equation 1) as a third 
independent calibration of the method.  Column 8 of Table \ref{tab:R23met} presents 
values for the R$_{23}$ abundances computed using the Tremonti \etal relation 
(hereafter referred to as the SDSS relation).  In the following sections, we utilize 
all three of these R$_{23}$ relations to calibrate our coarse abundance method and 
further explore the impact of different R$_{23}$ calibrations on the derived L-Z relation.

\begin{figure*}[htp]
\epsfxsize=5.0in
\epsscale{0.8}
\plotone{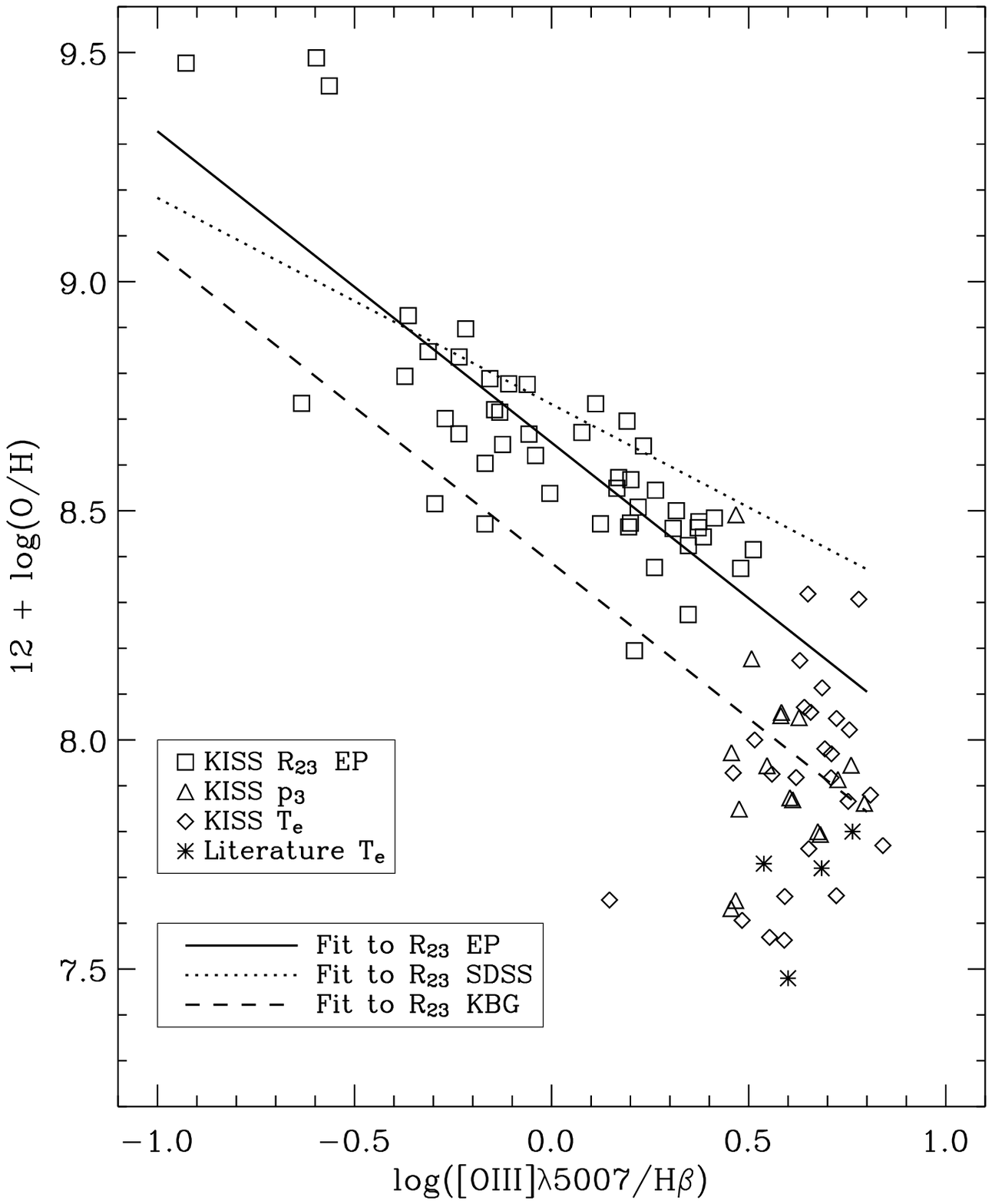}
\vskip -0.01in
\figcaption{This figure relates the [\ion{O}{3}]$\lambda$5007/H$\beta$ line ratio 
to metallicity. Abundances calculated using the EP $R_{23}$ fit are shown as squares, 
while $p_3$ abundances are shown as triangles.  Abundances calculated with the 
T$_3$ method are shown as diamonds.  T$_e$ abundance data taken from Izotov \etal 
(1997) are shown as asterisks.  We fit a line to the high metallicity branch (solid 
line). In addition, we plot the linear fits to the abundance data computed using 
the SDSS and KBG $R_{23}$ calibrations (dotted line and dashed line, respectively).  
See text for details.
\label{fig:o3met}}
\end{figure*}

\subsection{Coarse Metal Abundance Estimates}

Using the $T_e$, $p3$ and $R_{23}$ metal abundance estimates described above,
we calibrate empirical relations between two sets of strong emission-line
ratios ([\ion{N}{2}]$\lambda$6583/H$\alpha$, [\ion{O}{3}]$\lambda$5007/H$\beta$)
and metallicity.  We then estimate metal abundances for the large sample of 
KISS galaxies that have high quality follow-up spectra.  Figure \ref{fig:n2met} shows 
how the [\ion{N}{2}]$\lambda$6583/H$\alpha$ line ratio varies with metallicity, while
Figure \ref{fig:o3met} shows the same for the [\ion{O}{3}]$\lambda$5007/H$\beta$
line ratio.  For each plot, the diamonds are $T_e$ abundance measurements,
triangles are $p3$ abundances, and squares are $R_{23}$ measurements.  For the
latter we use the abundances computed using the EP R$_{23}$ 
calibration for the purpose of illustration.  Asterisks represent objects taken from Izotov \etal
(1997).  These objects are included to increase the number of low-metallicity galaxies 
available, in order to better constrain the resulting fit (as in Paper I). 

\begin{figure*}[htp]
\epsfxsize=5.0in
\epsscale{0.8}
\plotone{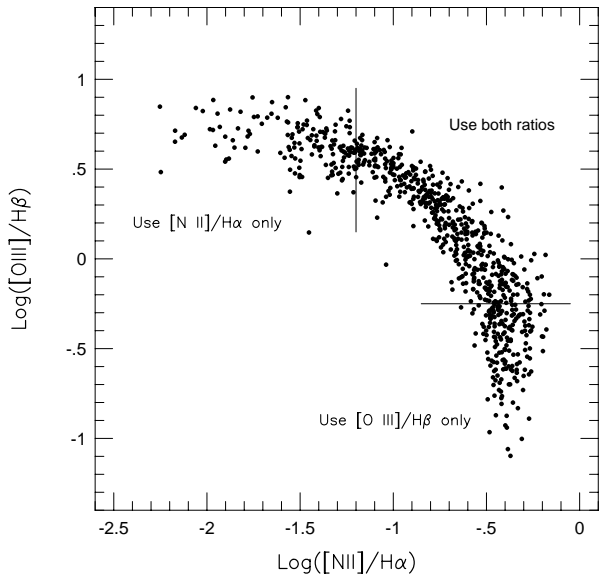}
\vskip -0.2in
\figcaption{A traditional emission-line diagnostic diagram plotting
[\ion{O}{3}]/H$\beta$ {\it vs.} [\ion{N}{2}]/H$\alpha$.  Only KISS star-forming
galaxies for which a coarse abundance estimate has been derived are included.  
The three regimes described in the text for specifying how the coarse abundances 
are computed are indicated.
\label{fig:diagplot}}
\end{figure*}

Our coarse abundance relations are determined by fitting low-order polynomials
to the data shown in Figures \ref{fig:n2met} and \ref{fig:o3met}.  We carry out the
fitting process separately for the three choices of the $R_{23}$ relation described
above.  For the [\ion{N}{2}]/H$\alpha$ -- metallicity data we fit a second order polynomial
to those objects with log([\ion{N}{2}]$\lambda$6583/H$\alpha$) $<-0.45$, where the line 
ratio is well behaved with respect to metallicity.  The fit is of the form
\begin{eqnarray}
12 + log(O/H) & = & A + B\cdot N + C\cdot N^2, \label{eqn:n2met} \\
N & = & log ([N II]\lambda6583/H\alpha). \nonumber
\end{eqnarray}
The polynomial coefficients and their formal uncertainties are given in Table \ref{tab:crsmet}.  
Also listed in the
table are the RMS scatters about the fits.  The latter lie in the range 0.13 to 0.16 dex,
and are adopted as the formal uncertainties in the abundance estimates obtained from 
this equation.  Note that the objects with high abundances (12 + log(O/H) $>$ 9.0) are 
excluded from the fit.  These are objects found in the lower right portion of the
diagnostic diagram shown in Figure~\ref{fig:diagplot}, where the [\ion{N}{2}]/H$\alpha$ 
ratio is no longer sensitive to changes in metallicity.  

We plot the three fits in
Figure \ref{fig:n2met}.  The solid line is the fit to the abundances derived with the EP
$R_{23}$ relation (corresponding to the data plotted in the figure), while the dashed
and dotted lines show the fits to the KBG and SDSS relations (data points not 
plotted).   All three fits agree well with each other on the left side of the diagram, 
where the [\ion{N}{2}]/H$\alpha$ ratio is most sensitive to metal abundance (see
below), but diverge on the right side due to the different $R_{23}$ abundance
determinations.  The KBG $R_{23}$  abundances are systematically lower than the
 EP ones by 0.26 dex, leading to the flatter  [\ion{N}{2}]/H$\alpha$ -- metallicity relation
 seen in the figure.   One the other hand, the SDSS fit agrees more closely with
 the EP relation, being only slightly steeper at the high-abundance end.

For the [\ion{O}{3}]/H$\beta$ -- metallicity data  we fit a linear relation to the objects located 
on the upper-metallicity branch of the  $R_{23}$ relation (log([\ion{N}{2}]$\lambda$6583/H$\alpha$) 
$>-1.2$).  The fit is of the form
\begin{eqnarray}
\label{eqn:o3met1}
12 + log(O/H) & = & A + B\cdot O,  \\
O &=& log([O III]\lambda5007/H\beta). \nonumber
\end{eqnarray}
The three different data sets are fitted separately, and the resulting polynomial coefficients 
are given in Table \ref{tab:crsmet}.   In this case the scatter about the fits ranges between
0.11 and 0.14, comparable to the values found for the [\ion{N}{2}]/H$\alpha$ relation.
Figure \ref{fig:o3met} plots the three fits, using the same conventions employed in Figure
\ref{fig:n2met}.   Since the fits are only to the upper branch objects, they more directly
reflect the differences in the input $R_{23}$ relations.  The KBG fit is parallel to, but
offset from, the EP fit, while the SDSS fit predicts lower abundances than EP for
the most metal-rich galaxies (left side of diagram) and higher abundances than EP for
galaxies with intermediate abundances.

Armed with these relations, we  calculate metal abundances for the large sample of KISS 
galaxies with high-quality follow-up spectra in the following way.  For galaxies with 
log([\ion{N}{2}]$\lambda$6583/H$\alpha$) $<$ $-$1.2 we calculate the metallicity 
using only the nitrogen line ratio and Equation \ref{eqn:n2met}.  For galaxies 
with log([\ion{N}{2}]$\lambda$6583/H$\alpha$) $>$ $-$1.2 and 
log([\ion{O}{3}]$\lambda$5007/H$\beta$) $>$ $-$0.25 we calculate metallicities 
using both the nitrogen and oxygen line ratios and equations \ref{eqn:n2met} 
and \ref{eqn:o3met1}.  We take the average of the two results to produce a final 
abundance estimate.  For objects with  log([\ion{O}{3}]$\lambda$5007/H$\beta$) 
$<$ $-$0.25, we use only the oxygen line ratio and  Equation \ref{eqn:o3met1} to 
calculate metallicity.  In Figure~\ref{fig:diagplot} we plot a standard emission-line 
diagnostic diagram of [\ion{O}{3}]/H$\beta$ {\it vs.} [\ion{N}{2}]/H$\alpha$, and 
indicate the three regions specified above.  In this diagram, the metallicity of
the galaxies increases smoothly from the upper left to the lower right.  It should 
be clear from the diagram that the [\ion{O}{3}]/H$\beta$ becomes 
insensitive to changes in metallicity at small values of [\ion{N}{2}]/H$\alpha$,
while the [\ion{N}{2}]/H$\alpha$ ratio is not a good abundance indicator at small
values of [\ion{O}{3}]/H$\beta$.

The above relations update and replace the corresponding versions from Paper I.   
The fit to the EP oxygen-metallicity plot is very similar to that from our
previous paper, while the EP nitrogen-metallicity relation has changed slightly at the
low abundance end.  For objects with log([\ion{N}{2}]/H$\alpha$) = $-$2.0, our new relation 
leads to abundances that are 0.07 dex higher, while at log([\ion{N}{2}]/H$\alpha$) = $-$1.2,
the revised relation predicts metallicities that are 0.03 dex higher than the old relation.
Pettini \& Pagel (2004) carry out an analysis very similar to ours for the [\ion{N}{2}]/H$\alpha$ 
ratio, and arrive at a relation reminiscent of our equation \ref{eqn:n2met}.  Their relation 
predicts oxygen abundances that are roughly 0.1 dex higher than those provided by our 
equation \ref{eqn:n2met} (for all three R$_{23}$ calibrations) for log([\ion{N}{2}]/H$\alpha$) 
less than -1.0, and comparable to our EP values for higher [\ion{N}{2}]/H$\alpha$ ratios.

%************************************************************************

\section{Optical and NIR Metallicity-Luminosity Relations}

\subsection{Presentation of new L-Z relations}

Following the procedures described in the previous section,  we are able to
compute coarse oxygen abundances for 766 KISS star-forming galaxies.  This
represents a roughly 50\% increase over the number of KISS ELGs with coarse
abundances available for the study carried out in Paper I.  Only
galaxies whose follow-up spectra provide high-quality measurements for the
[\ion{O}{3}]/H$\beta$ and [\ion{N}{2}]/H$\alpha$ line ratios are used.  This entire 
dataset also possesses B and V photometry, and more than half were detected by 
2MASS and have NIR magnitudes.  In this section we present luminosity-metallicity 
relations for the KISS ELGs in five different photometric bands.

\begin{figure*}[htp]
\epsfxsize=5.0in
\epsscale{0.99}
\plotone{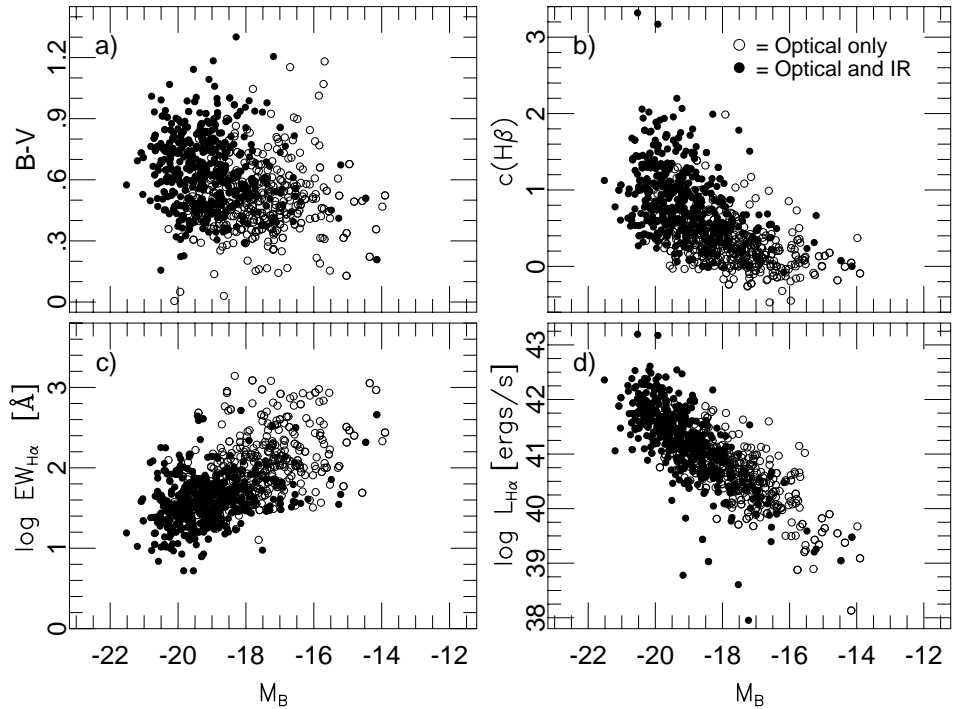}
\vskip -0.2in
\figcaption{Measured physical parameters for the KISS star-forming ELGs used
in this paper to construct our L-Z relations, plotted as a function of
B-band absolute magnitude: (a) B$-$V color; (b) the Balmer decrement reddening 
parameter c$_{H\beta}$; (c) the logarithm of the H$\alpha$ equivalent width 
(in \AA, measured from our follow-up spectra); and (d) the logarithm of the 
H$\alpha$ line luminosity (in ergs/s).  In all panels, filled symbols reflect
objects detected by 2MASS in the J band, while open symbols indicate those
that are not detected in the NIR.  Note the tendency for the lower luminosity 
galaxies to not be detected in the NIR. 
\label{fig:sample}}
\end{figure*}

In order to better illustrate the nature of our large sample of star-forming
ELGs, we display in Figure~\ref{fig:sample} plots of a number of measured physical
parameters.  The four panels plot the following parameters against B-band
absolute magnitude: (a) B$-$V color, (b) the Balmer decrement reddening 
parameter c$_{H\beta}$, (c) the logarithm of the H$\alpha$ equivalent width 
(in \AA, measured from our follow-up spectra), and (d) the logarithm of the 
H$\alpha$ line luminosity (in ergs/s).  In all panels, filled symbols reflect
objects detected by 2MASS in the J band, while open symbols indicate those
that are not detected in the NIR.  Ones sees expected/familiar trends in
all plots: the galaxies tend toward bluer colors, lower amounts of absorption,
higher equivalent width emission, and lower line luminosities in progressing
from higher to lower luminosities.  A key thing to note is the tendency for the 
lower luminosity galaxies to not be detected in the NIR: at luminosities above 
M$_B$ = $-$20.0, 96.8\% of the KISS ELGs are 2MASS detected, while for M$_B$ $>$ 
$-$18.0, only 20.4\% have measurable 2MASS fluxes.  A more complete description 
of the properties of the KISS ELGs can be found in L. Chomiuk \etal (2005, in 
preparation) and A. Jangren \etal (2005a, in preparation).

\begin{figure*}[htp]
\epsfxsize=6.0in
\epsscale{1.0}
\vskip -0.7in
\plotone{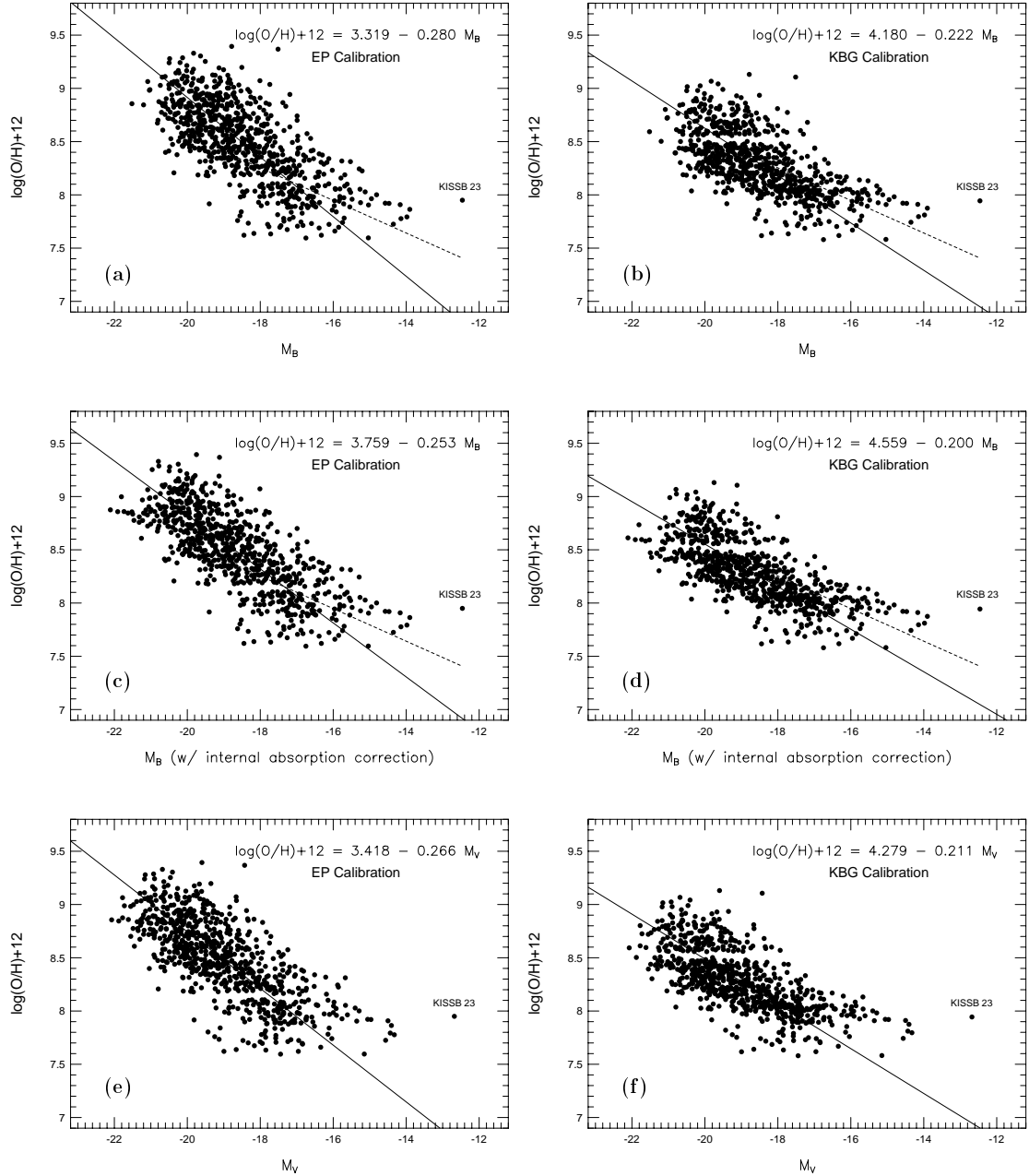}
\vskip -0.4in
\figcaption[lzOPT.ps]{Optical luminosity-metallicity relationships for the 
star-forming KISS ELGs (N=765).  Only galaxies with spectral qualities of 1 
or 2 are included.  The solid lines represent  bi-variate linear-least-squares fits to 
the data, while the dashed lines are the L-Z relation found by Skillman, Kennicutt
\& Hodge (1989).  Panels a) and b) show the B-band L-Z relations for the EP and
KBG R$_{23}$ calibrations respectively.  Panels c) and d) are the same pair, 
with the only difference being that an internal absorption correction has been applied.
Finally, panels e) and f) show the V-band L-Z relations for the EP and KBG relations.
\label{fig:lzOPT}}
\end{figure*}

\begin{figure*}[htp]
\epsfxsize=6.0in
\epsscale{1.0}
\vskip -0.7in
\plotone{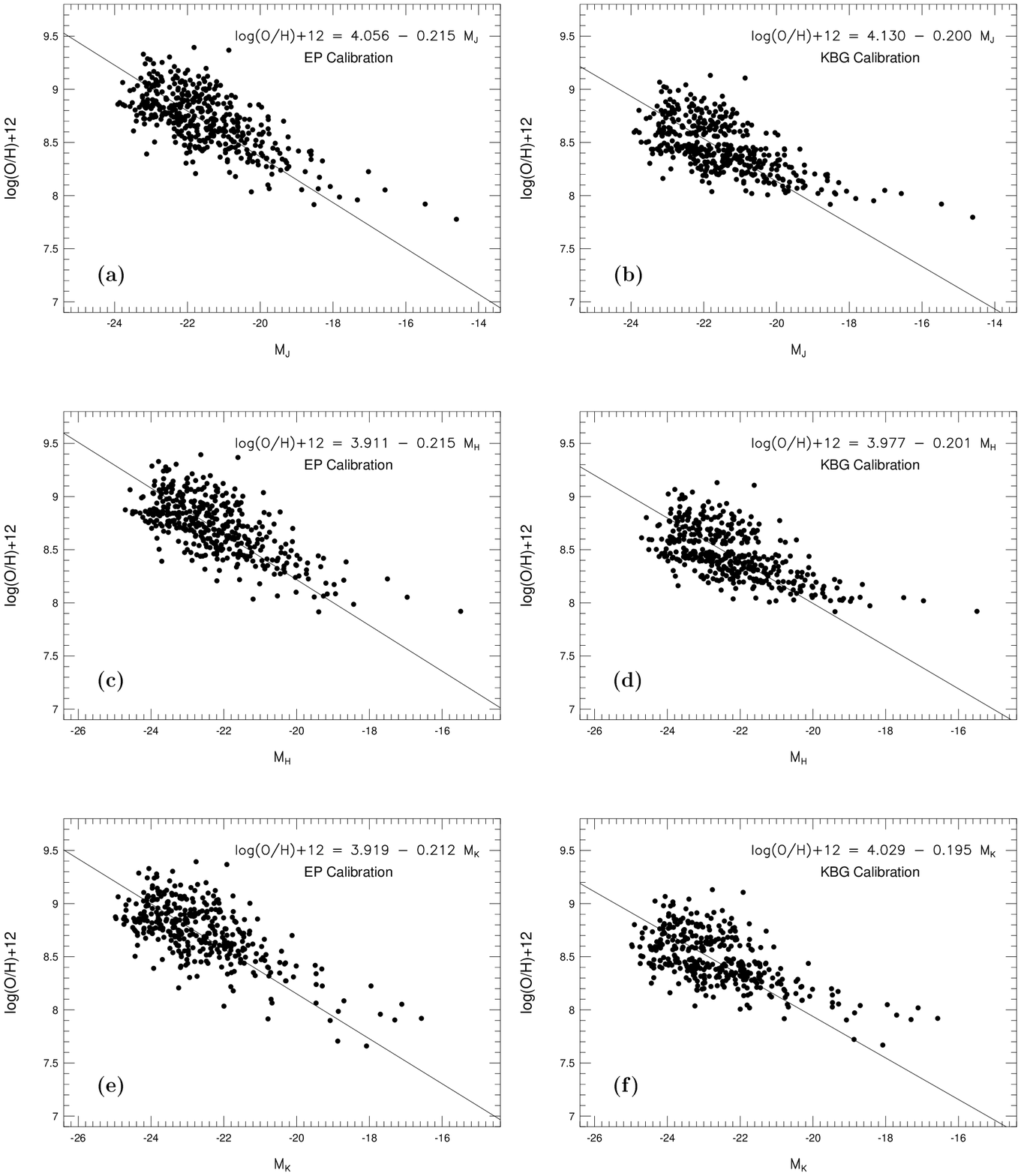}
\vskip -0.4in
\figcaption[lzNIR.ps]{Near infrared luminosity-metallicity relationships for the 
star-forming KISS ELGs.  Only galaxies with spectral qualities of 1 
or 2 are included.  The solid lines represent  bi-variate linear-least-squares fits to 
the data.  Panels a) and b) show the J-band L-Z relations for the EP and
KBG R$_{23}$ calibrations respectively (N=420).  Panels c) and d) present the
H-band relations for the EP and KBG relations (N=399).  
Finally, panels e) and f) show the K-band L-Z relations (N=370).
\label{fig:lzNIR}}
\end{figure*}

We present a series of L-Z plots in Figures~\ref{fig:lzOPT} and \ref{fig:lzNIR}.
In order to illustrate the effect of using different $R_{23}$ relations to compute the
abundances for the more luminous galaxies we plot side-by-side the EP and
KBG L-Z relations for each filter considered.  In all plots, the solid line represents 
a weighted bivariate linear-least-squares fit to the data (using the bisector method).
The uncertainties in the abundances are those specified in the previous
section.  For the uncertainties in the absolute magnitudes we adopt a 
conservative value of 0.3 mag.  This reflects mainly the errors in the distances.
The parameters of all of the linear-least-squares fits presented in this section 
are summarized in Table~\ref{tab:fits}.   The uncertainties listed in the table
for the slopes and intercepts are the formal errors obtained from the fitting
process.  The errors are quite small due to the large sample of galaxies
used.  They may not reflect the true uncertainties in the coefficients, however.
For example, the differences in the slopes of the direct and inverse fits are
substantially larger than the formal errors quoted in the table.  We do not
display the L-Z plots for the SDSS $R_{23}$ calibration here, but include the
results of the least-squares fits in Table~\ref{tab:fits}.  The appearance of
the SDSS L-Z plots are very similar to the EP versions shown.

In most previous work, the B-band has been the filter of choice for the
construction of L-Z relations.  This is presumably due to the wealth of
B-band photometry available for galaxies, but, as we shall see, it is perhaps a
poor choice due to the severe effects of internal absorption plus the impact
that extreme starbursts can have in the blue.
The B-band L-Z relations shown in Figure~\ref{fig:lzOPT}a,b include all 766 KISS
ELGs for which coarse abundances have been derived.  There is a clear trend 
of increasing metallicity with increasing luminosity, as expected.  The relations
are well defined between M$_B$ = $-$16 and $-$21, where the bulk of
the KISS sample is located.  A modest number of dwarf galaxies extend down to
M$_B$ = $-$14, but these all lie above the best-fit lines.  As suggested in
Paper I, this may be an indication that the L-Z relationship
flattens out at lower luminosities (i.e., the L-Z relation may not be strictly linear).
The slope of the fitted line is $-$0.280 for the EP relation, and $-$0.222 for the
KBG relation, both substantially steeper than B-band L-Z relations obtained in 
the past using primarily dwarf galaxy samples (e.g., Skillman, Kennicutt \& Hodge 
1989; Richer \& McCall 1995; Paper III).  The L-Z relation of Skillman \etal is shown in 
Figure~\ref{fig:lzOPT}a-d for comparison. The overall scatter in the O/H abundances 
about the fit line is 0.292 dex for EP and 0.245 dex for KBG.  This is larger than the 
formal errors in our coarse abundances, 
and suggests that there is substantial ``cosmic scatter" in the current relation.
We will return to each of these last three points in the following section.  

The location of the least luminous object in the figure,  KISSB 23 (a very nearby 
dwarf  ELG), is consistent with the idea that the L-Z relation may not be linear
down to the lowest luminosities.  It lies a full 1 dex above the extrapolation of the 
L-Z relation.   We have labeled it in Figure~\ref{fig:lzOPT}  because it
appears to deviate so strongly from the relationship defined by the rest of the
data set.  Part of the reason this object is so discrepant may have to do with
its evolutionary status.  KISSB 23 has an extremely low-excitation spectrum
compared to other dwarf ELGs.  It is located well below the main locus of
points in Figure~\ref{fig:diagplot} (log([\ion{N}{2}]/H$\alpha$) $\approx$ $-$1.45
and log([\ion{O}{3}]/H$\beta$) $\approx$ 0.15) and may be an aging
starburst.  Because of this, the coarse metal abundances based on the strong 
lines may be over-estimating the true metallicity.  We can confirm this suggestion,
since we have measured T$_e$ abundances for KISSB 23 (Paper II, Paper III); 
the latter study finds 12 + log(O/H) = 7.65, compared to 7.95
for our coarse abundance estimate (EP).   Galaxies like KISSB 23 show the limitations 
of the strong-lined method for estimating abundances (R$_{23}$, p3, our method).
Objects need to be near the peak of their star-formation episode (and on the 
main locus of points in line diagnostic diagrams like Figure~\ref{fig:diagplot}) in
order to generate accurate results from the strong-line abundance methods.
KISSB 23 is one of only a small handful of objects in our sample which may fall in 
the category of ``aging" starbursts.  Therefore, we have not included it in any of 
our fits.

As noted in Paper I, one reason the slope of the KISS
B-band L-Z relation is steeper than those of previous studies is likely due to
the fact that our sample contains a much higher proportion of luminous starburst
galaxies, which in turn likely have a large amount of internal absorption that
acts to decrease their observed luminosities.  If this increased absorption occurs 
systematically at the high-luminosity end of the L-Z relation, then it will naturally
lead to a steepening of the slope.  Following the ad-hoc procedure described in
Paper I, we can apply an internal absorption correction to the 
B-band absolute magnitudes and refit the data.  Our results are displayed in 
Figure~\ref{fig:lzOPT}c,d.  As expected, the slopes derived using the corrected 
luminosities are less steep (see Table \ref{tab:fits}).  Furthermore, the overall scatter 
in the relationships is reduced relative to the uncorrected fit.  
% We stress that the correction
%we use is almost certainly a lower limit to the actual amount of absorption, meaning 
%that the slope we find for the data in Figure~\ref{fig:lzOPT}c,d is most likely still
%steeper than the relationship one would obtain in the absence of absorption.

The reason we would like to arrive at an absorption-free form of the L-Z
relation is because such a relation should more nearly reflect the more 
fundamental relationship between metal abundance and mass.
An obvious way to approach directly the ideal of an absorption-free L-Z
relation is to measure the luminosity in longer wavelength photometric bands.
The KISS data set includes V-band photometry, which provides a modest
improvement over the B-band in this regard.  However, real gains can be
realized by moving into the NIR.  Figures~\ref{fig:lzOPT}e,f  and \ref{fig:lzNIR}  show
the KISS L-Z relations for the four photometric bands V, J, H, and K.  As
one would expect, the V-band relation (Figure~\ref{fig:lzOPT}e,f) is slightly less 
steep than for the uncorrected B-band (but more steep than the corrected 
B-band relation).  Again, the relation is well defined over more than five
magnitudes in absolute magnitude, and shows the same ``turn up" at the 
low-luminosity end exhibited by the B-band relation.  

The three sets of NIR L-Z plots (Figures~\ref{fig:lzNIR}) show a 
substantial change relative to the optical ones.  Unfortunately, the number 
of KISS star-forming galaxies that also possess 2MASS photometry is only
about half of the objects for which we have coarse abundance estimates.
In the J-band L-Z plot (Figure~\ref{fig:lzNIR}a,b), for example, only 420 (55.0\%)
of the galaxies in the coarse abundance sample are included.  Even lower
numbers apply to the H-band (399 galaxies, or 52.2\%) and K-band (370 galaxies,
or 48.4\%) data sets.  Additionally, the NIR data are preferentially lacking
at the lower luminosities, which motivated our efforts to obtain the additional
NIR data presented in \S 2.3.  Despite these constraints, the L-Z relations
in the NIR bands are well defined.

Two key results are readily apparent in the NIR L-Z plots.  First, all
three exhibit substantially lower slopes than are present in the relations
based on the optical data.  For all three, the slope is approximately $-$0.21 (EP)
or $-$0.20 (KBG).  Second, the RMS scatter in the L-Z relations is significantly lower 
for the NIR data compared to the optical data, at least for the EP and SDSS fits.  
Both results can be interpreted 
as being due, at least in part, to the reduced effects of dust absorption in the NIR.  
The shallower slopes come about as the strong, luminosity-dependent absorption 
in the B and V bands is reduced to much lower levels in the NIR bands.  
Further, since the amount of absorption varies tremendously from galaxy to
galaxy due to, for example, inclination effects, the scatter in the optical
L-Z relation is presumably elevated.  The lower scatter in the NIR relations
is therefore a natural consequence of ``removing" the variable amount of absorption.
These results are consistent with the findings of Masters, Giovanelli \& Haynes
(2003), who show that, on average, the amount of internal dust absorption
affecting disk galaxies is reduced to very low levels in the NIR compared
to the optical.  In particular, they suggest that the average level of absorption
in the K band is only 0.1 mag.  All appearances are that by producing a K-band
L-Z relation we have achieved our goal of nearly eliminating the effects
of absorption on the result.

\subsection{The Impact of Using Different R$_{23}$ Relations}

We have computed our coarse abundances using three independent calibrations for
the R$_{23}$ relation in order to evaluate the impact that the different versions 
might have on the derived L-Z relation.  The main results of this test can be seen
in Figures \ref{fig:lzOPT} \& \ref{fig:lzNIR} and Table \ref{tab:fits}, which clearly 
indicate that the choice of which R$_{23}$ relation to use for estimating the abundances 
of the high luminosity galaxies does have a significant impact on the slope of the
resulting luminosity-metallicity relation.

Examination of Figure \ref{fig:lzOPT} shows the the L-Z relations derived using the
KBG R$_{23}$ calibration are systematically shallower than the EP versions.  This
is to be expected, since the abundances of the most luminous galaxies in the sample
have abundances lower by 0.26 dex in the KBG relation.  The slope differences average
to $\sim$0.055 dex for the three different relations plotted.  Thus, if the KBG calibration is 
adopted over the EP calibration (or any of the other older determinations), the difference 
in the slope of the L-Z relation between our values based on EP and the ``classical"
Skillman, Kennicutt \& Hodge value ($-$0.15) is effectively reduced by half.
Interestingly, the slopes obtained using the SDSS calibration for the three optical
L-Z relations are only slightly lower than the EP versions.

The differences between the L-Z relations derived with the three R$_{23}$ calibrations
are much smaller when the NIR relations are considered.  The difference between the EP 
and KBG NIR slopes is only $\sim$0.015 dex, with the EP relation still being the steeper of 
the two.  The reason for the smaller slope difference in the NIR compared to the optical
is not immediately obvious, but presumably is rooted in the vagaries of the sample
differences between those KISS ELGs detected in the NIR and those that were not.
Perhaps even more surprising are the very shallow slopes obtained using the SDSS NIR
data.  We interpret this as being due to the flatter nature of the SDSS R$_{23}$
upper-branch relation (see Figure \ref{fig:o3met}).  We discuss this issue further
in the following section.  Why the {\it optical} SDSS L-Z fits agree so closely to the EP 
values, but then differ dramatically in the NIR, is again most likely an issue of the 
sample make up for the 2MASS-detected ELGs.

To summarize, we find significant differences in our derived L-Z relations depending on 
the choice of the R$_{23}$ (upper branch) relation employed.  At the moment there is
no clear-cut ``right" choice for which R$_{23}$ relation to use.  While the KBG version 
has the virtue of being based entirely on empirical data, the precise value of the derived
relation is heavily constrained by the handful of most metal-rich \ion{H}{2} regions
observed.  These observationally-derived abundances may well be uncertain, biasing the
resulting R$_{23}$ relation low.  We point out, for example, that the KBG relation
is 0.3 - 0.4 dex below two KISS ELGs with secure T$_e$-derived metallicities plotted in
Figure \ref{fig:o3met}.  On the other hand, the model-dependent calibrations used in all 
of the older R$_{23}$ relations (including EP) and by Tremonti \etal (2004) disagree with
each other at a significant level, and hence it is difficult to place full faith in them as
well.  It is not the purpose of this paper to try to make a firm assessment of which
R$_{23}$ relation is the best.  The answer to that question will likely require additional
work - both observational and theoretical.  Since our primary goal is to explore the 
nature of the L-Z relation, we are only concerned with the issue of how the different
R$_{23}$ relations impact the derived L-Z relation.  Our findings should serve as a
warning for future studies to carefully consider which R$_{23}$ relation to use and
to take care when comparing with previous studies.

%A key difference between the Edmunds \& Pagel 
%R$_{23}$ relation and some of the others is that the former is linear throughout the
%full range of observed R$_{23}$ values (for objects on the so-called upper branch),
%while those of McCall, Rybski \& Sheilds, McGaugh,  Zaritsky \etal, and Kobulnicky \etal 
%curve to lower abundance values for smaller values of R$_{23}$.  This curvature could
%in fact have a significant effect on our derived L-Z relation slopes (see \S 4).  

%************************************************************************

\section{Discussion}

Despite the warnings issued in the previous section regarding the variations seen in
our L-Z relations based on which R$_{23}$ calibration we use, one can still draw 
firm conclusions regarding the nature of the L-Z relation.  In the following discussion,
we attempt to minimize the impact that the different R$_{23}$ relations have by considering
the broader nature of the L-Z relations we have derived.  In essentially all cases, the
qualitative results do not depend on the specific choice for the R$_{23}$ calibration.

\subsection{Comparison with Previous Studies and the Slope of the B-band L-Z relation}

The results of Paper I and the current paper differ substantially
from the classical L-Z studies of Skillman \etal (1989) and Richer \& McCall (1995)
in the sense that the slope we obtain for the B-band luminosity-metallicity relation
is much steeper.   For example, Skillman \etal obtain a slope of $-$0.153 and a scatter 
of 0.16 for their sample of 19 quiescent irregular and dwarf irregular galaxies.  The recent
study by H. Lee \etal (2003) obtains nearly identical results.   In contrast, we find a slope of 
$-$0.280 (for the EP R$_{23}$ calibration, $-$0.222 for the KBG calibration) and a scatter 
of 0.29 dex (EP, 0.25 dex for KBG).  While part of the difference is due to the increased level 
of absorption in the higher luminosity galaxies in our sample, our attempt to correct for 
this only reduces the slope to $-$0.253 (EP, $-$0.200 for KBG).  We plot the L-Z relation 
of Skillman \etal as a dashed line in Figures~\ref{fig:lzOPT}a-d.

We do not believe that the slope differences listed above reflect any systematic 
problems with the respective studies.   In fact, when we use only the KISS galaxies 
with accurate T$_e$ abundances from Table~\ref{tab:temet} to construct an L-Z relation, 
we obtain a fit with precisely the same slope as Skillman \etal (see Paper III).  The 
galaxies in the latter sample cover the same absolute magnitude 
range as the Skillman \etal galaxies, and do not extend up into the higher luminosities 
present in our current sample.  Therefore, we believe that the slope differences between
these studies are real, and come about largely due to two factors: (1) The L-Z relation is 
not intrinsically linear over the full range of galaxian luminosities.  Therefore, studies
that cover different absolute magnitude ranges will arrive at different slopes.  Recent
studies involving large samples of galaxies have clearly revealed this non-linear nature
at the high-luminosity end
(e.g., Tremonti \etal 2004).  (2) The amount of intrinsic absorption in a galaxy scales 
with its metallicity: high abundance galaxies will tend to have a higher molecular gas 
and dust content.  Thus higher luminosity galaxies have luminosities that are systematically
underestimated, which compresses the absolute magnitude range at the luminous end
of the L-Z relation and artificially increases the slope.  Since the Skillman \etal and
Richer \& McCall samples are made up of lower luminosity galaxies, they occupy a
different portion of the L-Z diagram than does the bulk of the KISS sample.  They also
suffer less absorption than do the more luminous KISS ELGs.  The two effects
combine to produce a significantly shallower slope for these earlier studies, as well
as for the subsample of KISS ELGs that cover a similar absolute magnitude range.

Another possible factor that could potentially produce a difference between our results
and those of the previous studies mentioned above is the fact that we are using ``coarse"
abundances derived via strong emission lines, while the others are all using abundances
derived using the T$_e$ method.  However, this difference does not appear to have a
significant impact on the analysis.  We have T$_e$ abundances for 25 ELGs, taken mainly
from Papers II and III.  A comparison of the
T$_e$ and coarse abundances for these 25 sources shows that the two abundance estimates
agree within the expected uncertainties.  Specifically, the mean difference in log(O/H) between
the T$_e$ and coarse abundances is $-$0.009 dex, while the scatter about the mean is
0.163 dex (consistent with the uncertainty in the [\ion{N}{2}]/H$\alpha$ coarse abundance
method described in \S 3.3).  Thus, over the abundance range covered by our sample
of 25 T$_e$ objects (12+log(O/H) = 7.55 -- 8.30), our coarse abundance estimates are
in good agreement, albeit with a large scatter.

Further evidence for a steeper L-Z slope can be seen in Zaritsky, Kennicutt \& Huchra (1994).
Using spectra of disk \ion{H}{2} regions in a large sample of nearby spirals, they find that
an L-Z relation exists for gas-rich galaxies that covers approximately nine magnitudes in
M$_B$.  While they state that their sample maps directly onto the L-Z relation defined by
lower luminosity irregular galaxies, it is clear from their Figure 13 that the more luminous
spirals are located systematically above the relation for the irregulars, indicating that
a steeper slope is needed to fit the L-Z relation for the higher luminosity galaxies.

Garnett (2002) also considers the L-Z relation for a broad range of galaxian luminosities,
but finds a slope comparable to that obtained by Skillman \etal (1989),  Richer \& McCall (1995),
and H. Lee \etal (2003).  This appears to be in direct conflict with our findings, in the sense that it 
negates the explanation given above for the slope differences between our work and the dwarf-only
studies.  Garnett's sample includes many objects with M$_B$ brighter than $-$20.  The key
difference is that Garnett assigns a characteristic abundance for his larger spirals by
evaluating the observed abundance gradient for each galaxy and adopting the O/H value
measured at the half-light radius of the galaxy.  In contrast, we derive abundances measured 
in the {\bf central} star-forming complex of the KISS ELGs.  Since the abundance gradients in spirals 
are typically significant (Zaritsky, Kennicutt \& Huchra 1994; van Zee \etal 1998), the differences 
between our study and Garnett's can be understood in terms of {\it where} the metal
abundance is being measured in the galaxies.  For example, in the sample of van Zee \etal,
the abundance difference between the central value and the value measured at the half-light
radius averages 0.34 dex (with the central value being higher).  Adjusting the abundances of
Garnett's galaxies upward by 0.34 dex would clearly lead to a steeper L-Z slope.  The
question of where in the galaxy to measure the abundance is a valid one.  Certainly
one cannot fault Garnett for using the abundance measured part way out in the disk.
For the KISS ELGs, on the other hand, the choice is dictated by where the star-formation
event is occurring.  For luminous ELGs, the emission-line region is essentially always in
the central portion of the galaxy.  We would argue that for comparison with most modern
studies that utilize fiber-fed spectrographs, and for comparison with studies of higher
redshift galaxies, the use of centrally measured abundance values makes more sense.

Two recent studies have made use of much larger data sets to construct L-Z relations.
Lamareille \etal (2004)  utilize a sample of nearly 6400 galaxies from the 2dF Galaxy Redshift
Survey to determine a B-band luminosity-metallicity relation.   They select star-forming
galaxies from the 2dF database that have strong nebular emission lines from which the
O/H abundance can be derived using the R$_{23}$ method (using the McGaugh 1991 calibration).  
They obtain an L-Z slope very similar to our value ($-$0.274) and find a scatter in their 
relation of 0.27 dex.  Their sample contains a high percentage of luminous galaxies (M$_B$ $<$ 
$-$18): 92\% of their objects are in the upper branch of the R$_{23}$ relation.

Tremonti \etal (2004) use approximately 51,000 star-forming galaxies from the Sloan Digital
Sky Survey (SDSS) to create a pseudo-B-band L-Z relation as well as a mass-metallicity
relation.   As with the KISS and 2dF galaxies, they derive their metal abundances using
the strong nebular emission lines present in the spectra of their galaxies.  
A novel feature of the Tremonti \etal L-Z and mass-Z relations is that they flatten out at
the highest luminosities/masses.  This is readily understood in the context of the general
picture that the L-Z relation reflects mainly a galaxy's ability to retain its own SNe
ejecta, which depends primarily on the galaxian mass.  Above some critical
mass, essentially all of the metals produced in a galaxy will be retained.  Hence, one
would predict that for galaxian masses above this critical value, all galaxies will
have the same relative metal abundance, and the L-Z and mass-Z  relations should
flatten out.  This is exhibited dramatically with the SDSS data in Tremonti \etal!  Their
mass-Z relation begins to flatten for masses above a few times 10$^{10}$ M$_\odot$.
The flattening is less dramatic, but still evident, in their L-Z relation.  

Given the clear evidence for the flattening at the high luminosity/mass end of the 
observed L-Z and mass-Z relations derived with the SDSS data, one must ask the question 
why a similar trend is not seen in the KISS data (or, for that matter, in the 2dF data).  
Comparison of our Figure~\ref{fig:lzOPT}a with Figure 4 of Tremonti \etal suggests the 
answer: the SDSS database systematically
probes to higher luminosities than does KISS.  While KISS has few star-forming galaxies 
with M$_B$ $<$ $-$21, the SDSS sample of Tremonti \etal shows large numbers of
galaxies to M$_B$ = $-$22.5.   The differences between the two datasets are readily
understood as sample selection biases.  The spectroscopically observed portion of
SDSS is a magnitude-limited sample, and hence includes many luminous galaxies
at large redshifts.  KISS, on the other hand, is a line-selected survey that has an inherent
redshift limit of z = 0.095.  Hence KISS does not probe to the large distances (and large
volumes) that would allow it to detect large numbers of the relatively rare luminous galaxies.  
In addition, a second selection effect involves the ability to detect the line emission in KISS 
galaxies.  The absolute strength of the emission lines emitted by a star-forming event will
depend on several factors, but in general will scale with the size of the starburst.  Depending
on the size/luminosity of the host galaxy, that star-formation event will be detected as a
KISS galaxy only if there is sufficient contrast between the line and the underlying galaxian
continuum.  For extremely luminous host galaxies (M$_B$ $<$ $-$21), the underlying
galaxy light tends to swamp the line emission from the starburst, making them difficult to
detect.  This selection effect means that KISS under-represents star-formation in the very 
most luminous host galaxies, unless the star-formation event is extremely strong.  The two
effects (lower volume surveyed for luminous galaxies and the contrast effect) together
mean that KISS does not probe to the highest galaxian luminosities that SDSS reaches,
which explains, at least in part, why we do not see the flattening in the KISS L-Z relation.  
The flattening is only present in the most luminous galaxies, and these are rare in the KISS 
sample.

The derived SDSS L-Z relation has a slope of $-$0.185, which is considerably 
shallower than our result in the B-band but more consistent with the results 
of Skillman \etal (1989).  Examination of their Figure 2 suggests a reason 
for this discrepancy.  Tremonti \etal fit the entire dataset, including the 
portion with high luminosities where the correlation has begun to flatten out.  
It would appear that fitting only the galaxies with M$_B$ $>$  $-$21 would 
lead to a steeper slope.   This has been confirmed by C. Tremonti (private 
communication), who fit the SDSS data after applying a selection function 
that more nearly mimics the type of selection relevant for the KISS ELGs.  
The resulting slope of $-$0.23 is steeper than their published value and lies 
closer to the value we arrive at in the current study.

The flattening of the Tremonti \etal L-Z and mass-Z relations, as well as
their physical interpretation, needs also to be considered in the context of our
earlier discussion regarding the specifics of the metallicity calibration used.  
The abundance determination method employed by Tremonti \etal utilizes the strong-line
method but uses a more sophisticated scheme for deriving the metallicity.  This method is
model based, and yields an abundance calibration that tends to flatten dramatically at the
upper end of the R$_{23}$ relation (see their Figure 3).  While this flattening is reasonable,
if one were to to utilize a linear upper-branch R$_{23}$ relation (as we have for our EP and
KBG fits), then the flattening observed at the highest luminosities in the L-Z relation
essentially goes away (C. Tremonti, private communication).  While we are not 
suggesting that the observed flattening in the Tremonti \etal L-Z and mass-Z relations is 
fictitious, it is worth remembering that this result does depend strongly on the specifics of 
the strong-line abundance calibration used.

\subsection{The NIR L-Z relation}

One of the key results of the current paper is the derivation of NIR luminosity-metallicity
relations for our large sample of KISS star-forming galaxies.  Other groups have considered
the form of the L-Z relation in the NIR: Saviane \etal (2004) present an H-band relation for 
a small sample of Sculptor group galaxies, while Lilly, Carollo \& Stockton (2003) display
a J-band L-Z relation for their high redshift sample.  Neither group generates a fit to
their data, so we are unable to directly compare their relations with ours.  To our knowledge,
the NIR L-Z relations with fits shown in Figure~\ref{fig:lzNIR}  are the first ones
ever published for a local sample of galaxies.

As discussed in \S 4, the main motivation for constructing L-Z relations in the NIR is
to be able to avoid the effects of internal absorption on the high luminosity end of
the relation.  It is evident from the L-Z plots, as well as from the parameters of 
our linear-least-squares fits (Table~\ref{tab:fits}), that the slope of the L-Z relation 
decreases monotonically as the wavelength of the photometric band used to construct
it increases.  Furthermore, the RMS scatter about the fit is reduced in the
NIR relations.  The similarity of the slopes for the three NIR bands suggests that it doesn't
matter which band is used; the absorption effects are greatly diminished relative to
the optical bands for all three.

Because the sample of galaxies used to construct our NIR L-Z relations changes 
significantly relative to the one used to derive the optical relations (e.g., \S 4.1 and 
Figure~\ref{fig:sample}, we considered the possibility that the slope and RMS differences
seen might be due to sample differences rather than being a real physical effect. 
To test this, we created a new B-band L-Z relation using only the 420 galaxies included
in the J-band relation (Figure~\ref{fig:lzNIR}a,b).  The fit to this data set yielded a
slope of $-$0.307 $\pm$ 0.007 (EP calibration), slightly steeper than the value obtained 
with the full B-band sample ($-$0.280 $\pm$ 0.003).  Furthermore, the RMS scatter for
the new fit is 0.31 dex, compared to 0.29 dex for the fit to the full sample. {\it This result 
clearly shows that the differences in the L-Z relations between the optical and NIR are
NOT due to sample issues}.  In fact, one would have predicted just these results based on
the interpretations given in the previous subsection.  Since the NIR-detected sample is 
preferentially lacking in the lower luminosity galaxies, which are known to exhibit a 
shallower L-Z slope (see \S 5.1), the B-band L-Z slope for the J-band sample should be 
slightly steeper than the one for the full sample, as observed.  Further, the J-band-detected 
objects, being more luminous, will suffer more from variable amounts of internal absorption, 
leading to a larger RMS than when compared with the full sample.  Hence, we conclude that
the shallower slopes and reduced RMS scatters exhibited by our NIR L-Z relations are
real effects.

Another reason why the NIR L-Z relations are to be preferred over the optical ones is that 
variations in the NIR mass-to-light ratios (M/L) are substantially reduced relative to those in 
the optical (e.g., Bell \& de Jong 2001).  This is especially relevant if the underlying relation
is one between mass and metallicity, as was assumed in \S 1.  The primary effect of a smaller 
range of M/L values will be to further reduce the scatter in the L-Z relation by reducing the 
range in luminosities at a given mass.   This comes about because the effect of variations in
star-formation histories/stellar populations is less pronounced in the NIR than it is in the
optical.  This could be especially important for a sample of star-forming galaxies such as
KISS (although see \S 5.3).

The slope of the L-Z relation might also be expected to be shallower in the NIR simply
due to stellar population issues.  As shown by Bell \& de Jong (2001), the characteristic
colors of galaxies vary smoothly with M/L ratio, in the sense that lower M/L values are
bluer and higher M/L values are redder.   Color and M/L ratio also correlate with
luminosity, with bluer galaxies being less luminous and redder ones more luminous.
As an example, a galaxy with B$-$V = 0.3 and M$_B$ = $-$16.0 will have a B$-$K color
of $\sim$2.3 and hence M$_K$ = $-$18.3 (see Figure 9 of Bell \& de Jong 2001).   A
galaxy with B$-$V = 0.9 and M$_B$ = $-$21.0 will have  B$-$K $\approx$ 3.9 and 
M$_K$ = $-$24.9.  Hence, according to the models of Bell \& de Jong, one would
expect a shallower slope in the K-band L-Z relation since in this example $\Delta$M$_K$ 
= 6.6 when in the optical for the same two galaxies $\Delta$M$_B$ = 5.0.  Presumably,
both this effect and the lower amounts of absorption will both be acting simultaneously
to reduce the observed slope in the NIR L-Z relation.  It is not clear which effect will be
the dominant one (or whether they will be of roughly equal magnitude).   It is interesting
to note that for the KBG R$_{23}$ calibration, the L-Z slope is identical for the absorption-corrected
B-band relation and the NIR versions (Table \ref{tab:fits}).  This implies that all of the 
slope change between the B-band and NIR relations is due to absorption effects.   One 
the other hand, for both the EP and SDSS calibrations, the NIR L-Z slopes are significantly 
smaller than the absorption-corrected B-band values, suggesting that  a sizable portion of the 
slope difference is due to the correlation between color and luminosity.  Understanding the relative 
importance of the two effects will require additional work.  

In the absence of absorption (e.g., Masters \etal 2003), the slope of the L-Z relation is 
$-$0.21 $\pm$ 0.01 (EP calibration).  We suggest that this value is ``fundamental," in the 
sense that it reflects the change 
in abundance as a function of galaxian luminosity in the limit of zero absorption.  To the
extent that the L-Z relation is used to constrain models of galaxy evolution, the above
slope would be the preferred value to use.  While the ultimate goal would be to arrive
at a relationship between mass and metallicity, in the absence of the former
the NIR luminosity should be a robust substitute.  This is because the NIR luminosities
have a more direct correspondence to stellar mass than do optical luminosities.
The latter suffer from the effects of higher absorption, plus optical mass-to-light ratios
vary more than NIR ones due to stellar population differences.  Thus the NIR
luminosities are the preferred ones to use, and the NIR L-Z relations presented
here are likely to be close to the ideal or fundamental relationships between luminosity
and metallicity.

\subsection{Scatter in the L-Z relation and 2nd parameter effects}

Kobulnicky \etal (2003) searched for correlations between various parameters (color, 
emission-line equivalent width, effective radius) and residuals in their derived L-Z 
relationship in order to investigate whether there were any ``second-parameter effects"
that might be responsible for increased scatter in the L-Z relation.  They found no 
significant correlations.  In particular, they used the lack of correlation between H$\beta$
equivalent width and the L-Z residuals to argue that the presence of a strong starburst
in some galaxies does not add dramatically to the observed scatter in the L-Z relation.
While this result may appear to run counter to conventional wisdom, it is entirely
consistent with what is observed in the KISS sample of local star-forming galaxies.
For the luminous portion of the KISS sample (M$_B$ $<$ $-$20), even the most extreme 
starbursts rarely increase the total luminosity of the host galaxy by more than 0.2 
magnitude in B (A. Jangren \etal 2005a, in preparation).  In other words, we do not 
expect that the presence of starbursts 
in the KISS galaxies to cause a significant increase in the scatter in the L-Z relation at the
higher luminosities.  Since the Kobulnicky \etal  sample is confined to higher luminosities,
they naturally do not see a significant trend.  At lower luminosities the effect of a starburst on
the luminosity of the host galaxy, and hence its position in the L-Z diagram, can be more
significant (e.g., see Paper III).  However, even in the cases of extremely powerful
starbursts in low-luminosity hosts (a.k.a. blue compact dwarfs), the luminosity enhancement
due to the starburst is usually under 1.0 magnitude in B (e.g., Salzer \& Norton 1999).

We do see a significant reduction in the scatter in the L-Z relations derived using the
KISS galaxies when the NIR luminosities are used.  This suggests strongly that varying
amounts of internal absorption in galaxies add significantly to the scatter in the B-band
L-Z relation.   As seen in Table~\ref{tab:fits}, the RMS scatter in the KISS L-Z relation 
(EP calibration) drops from 0.29 dex for the uncorrected B-band relation, to 0.26 for the 
corrected B-band relation, to 0.22 for the J-band relation.   The fact that the scatter 
increases slightly in going from J to
H to K bands is likely due to the increasing photometric uncertainties present in the 2MASS
photometry in the redder bands (see \S 2.2).  As described above, the change in the derived 
slope for these fits is also consistent with a lowering  of the internal absorption on the
galaxian luminosities.  We stress that this absorption effect does not alter the derived
abundances, since the latter are corrected on a case-by-case basis for line-of-sight reddening 
based on their measured Balmer decrement.  This is strictly a luminosity effect, which is why
the scatter diminishes so dramatically between the B band and the NIR bands.

As discussed above, variations in the mass-to-light ratios of galaxies are expected to be
lower in the NIR bands than in the bluer optical bands.  It is likely that this effect also plays a 
role in reducing the scatter in the NIR L-Z relations.  It is not clear how one could disentangle
the two effects (absorption variations and M/L variations) with the present dataset.  

A key question is whether or not the observed scatter in the NIR bands (0.22 -- 0.23 dex)
represents the fundamental ``cosmic scatter" in the L-Z relation, or whether additional effects
are acting to inflate the observed values.  Since the formal errors in our abundance estimates
are of order 0.15 dex (and may be somewhat higher if systematic effects are important), we are 
probably not in a position to explore this issue further with the current dataset.  However, if we
adopt our formal abundance error as being a lower limit to the actual uncertainties in our
metallicity estimates,  we can set an upper limit on the cosmic scatter in the L-Z relation of 
$\sim$0.16 dex.  The lack of any strong correlations between the residuals in the J-band L-Z 
relation and H$\alpha$ equivalent width, c$_{H\beta}$, and B$-$V color suggest that the 
observed scatter is not caused primarily by the current star-formation processes.   Tremonti \etal
(2004) report that their metallicity residuals correlate fairly strongly with the central surface 
mass density of their sample galaxies.  This makes perfect sense, since if the L-Z relation is
caused by mass-dependent variations in the ability of different galaxies to retain their SNe
ejecta, then variations in the relation at a given mass/luminosity could naturally be caused
by variations in the {\it mass distributions} of the galaxies.  In other words, for galaxies of the
same mass, those with higher central mass densities should have modestly higher abundances.
The proper resolution of this issue will require additional work employing samples with more 
precise abundance estimates (e.g., Tremonti \etal 2004).

\subsection{High redshift L-Z relations}

Tracking the build-up of metals in galaxies over cosmic time time has recently become a 
realistic goal in the study of galaxy evolution.
Several groups have recently explored the change in the metal content of galaxies with
increasing redshift/look-back time (e.g., Lilly, Carollo \& Stockton 2003; Kobulnicky \etal 2003;
Maier, Meisenheimer \& Hippelein 2004, Liang \etal 2004).  With the establishment of a 
well-defined local L-Z relation, such as we have presented here,  it becomes possible 
to explore the evolution of the relation as a tool for probing the chemical (and possibly
star-formation) evolution history of galaxies.  However, the results to date are, to say the 
least, confused.  For example, the Maier, Meisenheimer \& Hippelein and Liang \etal studies 
both report that the galaxies in their sample exhibit metallicity increases of about a factor of 2 
between z $\approx$ 0.7 and today, while Kobulnicky \etal find more modest increases 
and Lilly, Carollo \& Stockton suggest that almost no metallicity enhancement has 
occurred over the same range of look-back times.

The task of making a self-consistent comparison between our new L-Z relations and the 
high redshift studies mentioned above is beyond the scope of the current paper.  While 
one might be tempted to simply compare the various results on face value, a more
careful approach is called for.   The comparison between our L-Z relation and the one
derived from SDSS data (Tremonti et al. 2004) serves as a clear example (see \S 5.1).
While a direct comparison of the L-Z fits to the two datasets suggest discrepant results,
a more careful analysis that treats the data in similar ways and accounts for sample
selection differences arrives at very similar L-Z fits.

The lessons we have learned from the comparison of purely local L-Z relations suggest 
some of the elements that are necessary for a study that seeks to measure the global 
chemical evolution in galaxies by contrasting L-Z relations at different epochs.  
These include: (1) {\it Estimating abundances for all datasets in identical
ways}.  This would require starting with the observed line ratios and processing the
high- and low-z samples following similar procedures.  Relevant issues here include 
accounting for internal absorption (difficult for high-z objects since H$\alpha$ is 
often redshifted out of the spectral bandpass, but crucial since abundance estimates
of high-z objects often rely on the [\ion{O}{2}]/H$\beta$ ratio which is extremely 
sensitive to absorption), and the consistent use of the same R23 (or equivalent abundance
estimator) calibration.  (2) {\it Proper corrections to rest-frame luminosities}.
While in principal this is very straight-forward (e.g., using standard k-corrections),
one needs to be concerned about issues like the variable amount of absorption
present in individual galaxies.  In practice, the best approach would be to use
multi-band imaging (which is often available for deep survey fields), fit the SED,
and then derive the appropriate rest-frame magnitude from the SED fit.  (3) {\it Use
of similar fitting methods for the L-Z relations}.  Different authors tend to use different 
methods for fitting their data.  To be assured that the fitting process does not introduce
differences, it is always advisable to use the same fitting software/method for
both high- and low-z L-Z relations.  (4) {\it Recognizing and accounting for sample
differences}.  As with most studies of extragalactic objects, one must always be
careful to avoid comparing ``apples and oranges."  Since high-z samples tend to
be dominated by higher luminosity galaxies, care must be taken to ensure that the
local comparison sample has a similar make up.

The prospect of using L-Z relations measured at a range of redshifts as a tool for
probing the chemical evolution of galaxies is an exciting one.  While only in its infancy,
this technique offers great promise for our understanding of galaxy evolution as well
as the star-formation history of the universe.   However, as outlined above, great care
is required to avoid drawing incorrect conclusions.

\section{Summary \& Conclusions}
We use the large sample of nearby (z $<$ 0.095) star-forming galaxies from the
KISS objective-prism survey to explore the relationship between metallicity and luminosity.  
The homogeneous nature of the KISS follow-up spectra, combined with the depth and uniformity 
of the sample make this an excellent dataset to use for this purpose.
We extend the analysis presented in Paper I by
(i) using an expanded sample of KISS star-forming galaxies with the necessary follow-up
spectra, (ii) re-determining the coarse abundance relations with a much larger sample
of objects with quality abundance estimates, (iii) exploring the effect of using different
R$_{23}$ calibrations on the resulting L-Z relation, and (iv) evaluating the L-Z relation in
five different photometric bands (BVJHK).

Follow-up spectra of a total of 766 KISS star-forming ELGs
enable us to estimate abundances via a coarse (or strong-lined) abundance
method.   These objects also have accurate B and V photometry.  By 
correlating the KISS catalog with the 2MASS database, we obtain NIR magnitudes for
a subset of these KISS ELGs: 420 with J-band photometry, 399 with H-band data, and
370 with K-band magnitudes.

We demonstrate that testing the effect of using different strong-line method 
calibrations on the resulting L-Z relations is important and necessary, and 
expand our analysis from Paper I to include a range of proposed R$_{23}$-method 
calibrations.  These result in significant variations in our derived L-Z relation 
fits.  While the range in the derived L-Z slopes is significant, especially in the 
optical, we do not believe that it changes any of our general conclusions.

We present L-Z relations for all five photometric bands.  In all cases the relations
are derived using data that cover more than six magnitudes, or a factor of 250 in 
luminosity.  We find that the slopes of the L-Z relations change
systematically from the shortest to the longest wavelengths, in
that the relation is steepest in the blue and most shallow in the near-IR.
The RMS scatter in our relations decrease dramatically between the blue and NIR.  
We interpret these effects as being due, in large part, to the changing influence
of internal absorption on the intrinsic brightnesses of the luminous portions of the
galaxy populations as a function of the observed wavelength.  In addition, changes
in the contributions of the various stellar populations present in these galaxies
(e.g., starburst population vs. older stars) to the various bandpasses likely 
contributes to the observed slope changes as well.

The slope of our derived B-band L-Z relation differs dramatically from those obtained
in some previous studies (e.g., Skillman \etal (1989);
Richer \& McCall 1995).  The main difference is that we obtain a much
steeper slope in the optical.  However, those studies considered only
low luminosity galaxies (typically M$_B$ $>$ $-$18), while the KISS sample 
ranges from M$_B$ = $-$14 to $-$21.5.   Recent studies that cover a similar luminosity
range also derive larger slopes (e.g., Lamareille \etal 2004).  While part of the difference 
between our results and previous work may be due to absorption
effects in the more luminous galaxies, it appears that the general L-Z
relation is steeper than the value obtained using only low-luminosity galaxies.
We suggest that the K-band slope of $-$0.21 represents a ``fundamental" form
of the L-Z relation.  The fact that the slope derived using low-luminosity galaxies
is less steep ($-$0.15) suggests that the L-Z relation is not linear over the full
range of galaxian luminosities.  The large scatter in our data, combined with the
relative paucity of galaxies with luminosities below M$_B$ = $-$16, makes it 
impossible for us to explore the non-linear nature of the L-Z relation with the
current dataset.  However, a clear hint of this nature can be seen in the L-Z
and mass-Z relations derived recently by Tremonti \etal using the extensive
SDSS database.

The determination of accurate L-Z relations provides fundamental input to our
understanding of galaxy evolution.  The fact that metal abundance
correlates with galaxian luminosity (and mass) over essentially the entire range
of the latter suggests that there are important constraints on the supernova
feedback processes occurring within galaxies.  Furthermore, the observed
dispersion in the L-Z relation sets limits on the range of variation of these
feedback mechanisms from galaxy to galaxy.  Together, they provide important
constraints to models of the chemical evolution of galaxies (e.g., Prantzos \& Boissier 
2000; Boissier \etal 2003;  Chiappini, Romano \& Matteucci 2003;  Rieschick \& Hensler 
2003; Qian \& Wasserburg 2004).  In addition, recent studies have started to reveal 
the utility of using the L-Z relation as a tool for measuring the evolution of metal 
abundances in galaxies as a function of look-back time.

%************************************************************************

\acknowledgements

We gratefully acknowledge financial support for the KISS project from an NSF 
Presidential Faculty Award to JJS (NSF-AST-9553020), as well as continued 
support for our ongoing follow-up spectroscopy campaign (NSF-AST-0071114).
JCL acknowledges financial support from NSF grants AST-9617826 and AST-0307386
as from a UA-NASA Space Grant Graduate Fellowship.  
We would also like to recognize and thank the members of the 2MASS project
for producing such an excellent, useful, and highly accessible data set for 
the community.  We are grateful for the assistance of Sun Mi Chung, Joseph
Martin, and Anthony Moeller who helped with the analysis of the 2MASS data.
We thank Don McCarthy for the use of his PISCES NIR camera, as well as George 
and Marcia Rieke for the use of their 256 $\times$ 256 NIR camera.  Our NIR 
observations would not have been possible without these instruments and the 
respective PI's technical support during our runs.
We thank the many KISS team members who have participated in the spectroscopic 
follow-up observations during the past several years, particularly Caryl 
Gronwall, Drew Phillips, Gary Wegner, Jessica Werk, Laura Chomiuk, Kerrie 
McKinstry, Robin Ciardullo, Jeffrey Van Duyne and Vicki Sarajedini.
We benefited greatly from helpful conversations with Christy Tremonti, Don
Garnett, and Rob Kennicutt.
Many useful suggestions by the referee, Chip Kobulnicky, helped to improve 
the presentation of this paper.  
Finally, we wish to thank the support staffs of Kitt Peak National Observatory,
Lick Observatory, the Hobbey-Eberly Telescope,  MDM Observatory, and Apache
Point Observatory for their excellent assistance in obtaining the spectroscopic
observations that made this work possible.

%********************************* REFERENCES***************************

%clearpage

% Now comes the reference list.  In this document, we used \cite to call
% out citations, so we must use \bibitem in the reference list, which
% means we use the LaTeX thebibliography environment.  Please note that
% \begin{thebibliography} is followed by a null argument.  If you forget
% this, mayhem ensues, and LaTeX will say "Perhaps a missing item?" when
% you run it.  Do not call us, do not send mail when this happens.  Put
% the silly {} after the \begin{thebibliography}.
%
% Each reference has a \bibitem command to define the citation format
% to be placed in the text (in []) and the symbolic tag used for 
% cross referencing (in {}).
%
% See sample1.tex, or the AASTeX guide, for an alternative to the \cite-
% \bibitem command.

%********************************* TABLE 1 ************************************
\clearpage

%\tabletypesize{\small}
%\renewcommand{\arraystretch}{.6}     

\begin{deluxetable}{rccccc}
\tablewidth{0pt }
\tablecaption{New H \& K$_S$ Band Photometry \label{tab:nirphot}}
\tablehead{
\colhead{KISSR} & \colhead{KISSB} & \colhead{H} &
\colhead{$\sigma_H$} & \colhead{K$_S$}  & \colhead{$\sigma_K$} \\
\colhead{(1)}& \colhead{(2)}& \colhead{(3)}& 
\colhead{(4)}& \colhead{(5)}& \colhead{(6)} 
}
\startdata
 286   & 133 &\nodata&\nodata&  14.60 &  0.06   \\
 310   & 136 &\nodata&\nodata&  16.92 &  0.07   \\
 311   & 137 &\nodata&\nodata&  16.49 &  0.08   \\
 314   & 138 &12.98  & 0.19  &\nodata &\nodata  \\
 396   & 145 &\nodata&\nodata&  15.16 &  0.06   \\
 505   & 161 &14.30  & 0.30  &\nodata &\nodata  \\
 666   & 186 &\nodata&\nodata&  17.59 &  0.11   \\
 678   & &16.24  & 0.20  &  15.34 &  0.06   \\
 742   & 192 &16.53  & 0.38  &\nodata &\nodata  \\
 998   & &16.07  & 0.39  &\nodata &\nodata  \\
1014   & &16.95  & 0.27  &\nodata &\nodata  \\
1091   & &16.45  & 0.45  &\nodata &\nodata  \\ 
1123   & 222 &15.33  & 0.16  &\nodata &\nodata  \\
\enddata
\end{deluxetable}

%********************************* TABLE 2 ************************************

\tabletypesize{\small}
\renewcommand{\arraystretch}{.6}     

\begin{deluxetable}{cccccc}
\tablewidth{0pt }
\tabletypesize{\footnotesize}
\tablecaption{Metal abundances from the $T_e$ method. \label{tab:temet}}
\tablehead{
\colhead{KISSR} & \colhead{KISSB} & \colhead{log([\ion{N}{2}]/H$\alpha$) \tablenotemark{a}} &
\colhead{log([\ion{O}{2}]/H$\beta$) \tablenotemark{a}} & \colhead{log([\ion{O}{3}]/H$\beta$) 
\tablenotemark{a,b}} & \colhead{12 + log(O/H)}\\
\colhead{(1)}& \colhead{(2)}& \colhead{(3)}& 
\colhead{(4)}& \colhead{(5)}& \colhead{(6)}
}
\startdata
   ...& 23 &   -1.450&     0.393&     0.147&     7.651\\
   ...& 61 &   -2.160&     0.142&     0.652&     7.763\\
   ...& 71 &   -1.989&    -0.087&     0.722&     7.660\\
   ...& 86 &   -1.409&     0.332&     0.657&     8.060\\
    49& 94 &   -1.099&     0.537&     0.515&     8.000\\
    73& 98 &   -1.519&     0.417&     0.559&     7.925\\
    85& ...&   -2.444&     0.069&     0.483&     7.606\\
    87& 104&   -1.101&     0.278&     0.650&     8.318\\
   116& 107&   -1.196&     0.304&     0.641&     8.071\\
   286& 133&   -1.269&     0.312&     0.630&     8.174\\
   310& 136&   -1.940&    -0.020&     0.809&     7.880\\
   311& 137&   -1.539&     0.378&     0.693&     7.981\\
   396& 145&   -1.531&     0.396&     0.620&     7.918\\
   ...& 171&   -1.416&     0.263&     0.686&     8.114\\
   ...& 175&   -1.514&     0.253&     0.755&     8.022\\
   666& 186&   -2.166&    -0.863&     0.840&     7.770\\
   675& 187&   -1.706&     0.103&     0.752&     7.866\\
   814& 199&   -1.314&     0.259&     0.710&     7.970\\
   818& 200&   -1.280&     0.459&     0.779&     8.307\\
  1013& 211&   -1.259&     0.320&     0.592&     7.658\\
  1194& ...&   -1.506&     0.311&     0.709&     7.918\\
  1490& ...&   -1.606&     0.179&     0.590&     7.563\\
  1752& ...&   -1.897&     0.038&     0.553&     7.569\\
  1778& ...&   -1.295&     0.447&     0.461&     7.928\\
  1845& ...&   -1.429&     0.385&     0.723&     8.047\\
\enddata

\tablenotetext{a}{All line ratios are corrected for internal absorption using the measured
value of c$_{H\beta}$.  No corrections for underlying Balmer absorption have been applied.}
\tablenotetext{b}{[\ion{O}{3}]$\lambda\lambda$5007/H$\beta$}
\end{deluxetable}

%********************************* TABLE 3 ************************************

%\clearpage

\tabletypesize{\small}
\renewcommand{\arraystretch}{.6}     

\begin{deluxetable}{cccccc}
\tablecaption{Metal abundances from the $p_{3}$ method. \label{tab:p3met}}
\tablewidth{0pt }
\tabletypesize{\footnotesize}
\tablehead{
\colhead{KISSR} & \colhead{KISSB} & \colhead{log([\ion{N}{2}]/H$\alpha$) \tablenotemark{a}} &
\colhead{log([\ion{O}{2}]/H$\beta$) \tablenotemark{a}} & \colhead{log([\ion{O}{3}]/H$\beta$) \tablenotemark{a,b}} & 
\colhead{12 + log(O/H)}\\
\colhead{(1)}& \colhead{(2)}& \colhead{(3)}& 
\colhead{(4)}& \colhead{(5)}& \colhead{(6)}
}
\startdata
 ... &    47&    -1.782&     0.148&     0.682&     7.794\\
   96&   105&    -1.359&     0.280&     0.726&     7.914\\
   97&  ... &    -1.337&     0.493&     0.628&     8.049\\
  105&  ... &    -1.502&     0.166&     0.467&     7.649\\
  120&  ... &    -1.367&     0.167&     0.675&     7.800\\
  223&   127&    -1.324&     0.513&     0.583&     8.060\\
  272&  ... &    -1.528&     0.359&     0.476&     7.849\\
  343&   140&    -1.396&     0.331&     0.605&     7.874\\
  404&  ... &    -1.489&     0.459&     0.456&     7.972\\
  405&   147&    -1.503&     0.154&     0.455&     7.632\\
  528&   164&    -1.437&     0.507&     0.582&     8.053\\
  715&   190&    -1.728&     0.061&     0.794&     7.862\\
  785&   194&    -1.565&     0.287&     0.760&     7.945\\
  803&  ... &    -1.505&     0.600&     0.507&     8.177\\
 1021&  ... &    -1.325&     0.775&     0.468&     8.491\\
 1177&  ... &    -1.498&     0.423&     0.547&     7.944\\
 1794&  ... &    -1.492&     0.322&     0.612&     7.869\\
\enddata
\tablenotetext{a}{All line ratios are corrected for internal absorption using the measured
value of c$_{H\beta}$.  No corrections for underlying Balmer absorption have been applied.}
\tablenotetext{b}{[\ion{O}{3}]$\lambda$5007/H$\beta$}
\end{deluxetable}

%********************************* TABLE 4 ************************************

%\clearpage

\tabletypesize{\small}
\renewcommand{\arraystretch}{.6}     

\begin{deluxetable}{cccccccc}
\tablecaption{Metal abundances from the $R_{23}$ method. \label{tab:R23met}}
\tablewidth{0pt }
\tabletypesize{\footnotesize}
\tablehead{
\colhead{KISSR} & \colhead{KISSB} & \colhead{log([\ion{N}{2}]/H$\alpha$) \tablenotemark{a}} &
\colhead{log([\ion{O}{2}]/H$\beta$) \tablenotemark{a}} & \colhead{log([\ion{O}{3}]/H$\beta$) \tablenotemark{a,b}} & &
\colhead{12 + log(O/H)}\\
&&&&& EP & KBG & SDSS\\
\colhead{(1)}& \colhead{(2)}& \colhead{(3)}& \colhead{(4)}& 
\colhead{(5)}& \colhead{(6)}& \colhead{(7)}& \colhead{(8)}
}
\startdata
     1&    ... &    -0.683&     0.651&     0.124&     8.472&     8.209&     8.607\\
     3&    ... &    -0.571&     0.729&    -0.169&     8.471&     8.209&     8.606\\
    68&     97 &    -0.787&     0.426&     0.413&     8.484&     8.222&     8.619\\
    91&    ... &    -0.750&     0.549&     0.166&     8.549&     8.287&     8.679\\
   109&    ... &    -0.527&     0.506&     0.201&     8.568&     8.305&     8.695\\
   110&    ... &    -0.648&     0.434&    -0.109&     8.777&     8.514&     8.857\\
   117&    ... &    -0.963&     0.484&     0.374&     8.476&     8.214&     8.611\\
   124&    ... &    -0.859&     0.506&     0.372&     8.463&     8.201&     8.598\\
   133&    ... &    -0.576&     0.890&     0.210&     8.195&     7.933&     8.293\\
   145&    ... &    -0.368&     0.420&    -0.235&     8.836&     8.573&     8.895\\
   185&    ... &    -0.533&     0.355&    -0.218&     8.897&     8.635&     8.932\\
   257&    ... &    -0.811&     0.562&     0.308&     8.461&     8.199&     8.597\\
   265&    ... &    -0.626&     0.489&     0.264&     8.545&     8.283&     8.675\\
   292&    135 &    -0.736&     0.522&     0.480&     8.374&     8.112&     8.507\\
   317&    ... &    -0.965&     0.404&     0.512&     8.415&     8.153&     8.550\\
   322&    ... &    -0.413&    -0.043&    -0.564&     9.427&     9.164&     9.149\\
   326&    ... &    -0.613&     0.631&    -0.005&     8.538&     8.276&     8.669\\
   327&    ... &    -0.717&     0.580&     0.347&     8.424&     8.162&     8.559\\
   333&    ... &    -0.325&     0.430&    -0.313&     8.847&     8.585&     8.903\\
   335&    ... &    -0.384&     0.386&     0.233&     8.641&     8.379&     8.757\\
   368&    ... &    -0.483&     0.711&    -0.296&     8.515&     8.253&     8.648\\
   398&    ... &    -0.427&     0.501&    -0.145&     8.721&     8.458&     8.817\\
   400&    ... &    -0.430&     0.371&    -0.364&     8.926&     8.663&     8.949\\
   515&    ... &    -0.528&     0.524&    -0.058&     8.667&     8.405&     8.777\\
   520&    ... &    -0.387&     0.440&    -0.157&     8.788&     8.526&     8.865\\
   577&    ... &    -0.422&     0.572&    -0.235&     8.668&     8.405&     8.778\\
   582&    ... &    -0.968&     0.759&     0.347&     8.273&     8.011&     8.392\\
   595&    ... &    -0.399&     0.005&    -0.928&     9.477&     9.214&     9.163\\
   610&    ... &    -0.662&     0.461&     0.077&     8.671&     8.408&     8.780\\
   643&    182 &    -0.856&     0.505&     0.317&     8.500&     8.238&     8.634\\
   725&    ... &    -0.634&     0.348&     0.191&     8.696&     8.433&     8.799\\
   830&    ... &    -0.474&     0.690&     0.261&     8.376&     8.114&     8.509\\
   833&    ... &    -0.542&     0.566&    -0.124&     8.645&     8.382&     8.760\\
   894&    ... &    -0.493&     0.614&    -0.169&     8.603&     8.341&     8.726\\
   910&    ... &    -0.566&     0.416&    -0.062&     8.776&     8.513&     8.856\\
  1016&    ... &    -0.500&     0.502&    -0.131&     8.715&     8.453&     8.813\\
  1028&    ... &    -0.362&    -0.092&    -0.597&     9.488&     9.225&     9.165\\
  1032&    ... &    -0.723&     0.520&     0.170&     8.572&     8.310&     8.699\\
  1055&    ... &    -0.466&     0.550&    -0.270&     8.701&     8.439&     8.803\\
  1062&    ... &    -0.640&     0.565&     0.220&     8.508&     8.246&     8.641\\
  1402&    ... &    -0.337&     0.571&    -0.634&     8.734&     8.472&     8.827\\
  1416&    ... &    -0.719&     0.615&     0.200&     8.473&     8.211&     8.608\\
  1424&    ... &    -0.292&     0.490&    -0.372&     8.793&     8.531&     8.868\\
  1537&    ... &    -0.701&     0.522&     0.385&     8.443&     8.181&     8.578\\
  1885&    ... &    -0.727&     0.364&     0.112&     8.733&     8.471&     8.827\\
  1940&    ... &    -0.767&     0.626&     0.196&     8.465&     8.203&     8.600\\
  2021&    ... &    -0.592&     0.564&    -0.041&     8.621&     8.358&     8.740\\
\enddata
\tablenotetext{a}{All line ratios are corrected for internal absorption using the measured
value of c$_{H\beta}$.  No corrections for underlying Balmer absorption have been applied.}
\tablenotetext{b}{[\ion{O}{3}]$\lambda$5007/H$\beta$}
\end{deluxetable}

%********************************* TABLE 5 ************************************

%\clearpage

%\tabletypesize{\small}
%\renewcommand{\arraystretch}{.6}     

\begin{deluxetable}{lcccc}
\tablecaption{Coarse Metallicity Relations \label{tab:crsmet}}
\tablewidth{0pt }
\tablehead{
\colhead{R$_{23}$ Relation} & \colhead{A} & \colhead{B} & \colhead{C} & \colhead{RMS}\\
& & & & \colhead{Scatter}\\
\colhead{(1)}& \colhead{(2)}& \colhead{(3)}& \colhead{(4)}& \colhead{(5)}
}
\startdata
& & \\
&& [\ion{N}{2}]/H$\alpha$ Relations\\
& & \\
EP      & 9.170 $\pm$ 0.070 & 1.094 $\pm$ 0.098 & \ \ 0.175 $\pm$ 0.033 & 0.136\\
KBG   & 8.578 $\pm$ 0.070 & 0.423 $\pm$ 0.098 & $-$0.009 $\pm$  0.033 & 0.156\\
SDSS & 9.433 $\pm$ 0.070 & 1.391 $\pm$ 0.098 & \ \ 0.256 $\pm$ 0.033 & 0.130\\
\hline
& & \\
&& [\ion{O}{3}]/H$\beta$ Relations\\
& & \\
EP      & 8.649 $\pm$ 0.022 & $-$0.679 $\pm$ 0.068 & & 0.144\\
KBG   & 8.386 $\pm$ 0.022 & $-$0.679 $\pm$ 0.068 & & 0.144\\
SDSS & 8.732 $\pm$ 0.022 & $-$0.450 $\pm$ 0.068 & & 0.112\\
\enddata
\end{deluxetable}

%********************************* TABLE 6 ************************************

%\clearpage

%\tabletypesize{\small}
%\renewcommand{\arraystretch}{.6}     

\begin{deluxetable}{lcccccc}
\tablecaption{Results of Linear Fits to the L-Z Relations \label{tab:fits}}
\tablewidth{0pt }
\tablehead{
\colhead{Filter} & \colhead{Number} & \colhead{Slope} &
\colhead{Slope} & \colhead{Intercept}  & \colhead{Intercept} & \colhead{RMS}\\
& & & \colhead{Uncertainty} & & \colhead{Uncertainty} & \colhead{Scatter}\\
\colhead{(1)}& \colhead{(2)}& \colhead{(3)}& 
\colhead{(4)}& \colhead{(5)}& \colhead{(6)} & \colhead{(7)}
}
\startdata

&&&EP $R_{23}$ Relation \\
B & 765 & $-$0.280 & 0.003 & 3.32 & 0.06 & 0.292\\
B$_o$ & 765 & $-$0.253 & 0.002 & 3.76 & 0.05 & 0.261\\
V & 765 & $-$0.266 & 0.002 & 3.42 & 0.05 & 0.274\\
J & 420 & $-$0.215 & 0.003 & 4.06 & 0.09 & 0.223\\
H & 399 & $-$0.215 & 0.003 & 3.91 & 0.09 & 0.225\\
K & 370 & $-$0.212 & 0.003 & 3.92 & 0.09 & 0.235\\
\hline
&&& KBG $R_{23}$ Relation \\
B & 765 & $-$0.222 & 0.003 & 4.18 & 0.06 & 0.245\\
B$_o$ & 765 & $-$0.200 & 0.002 & 4.56 & 0.05 & 0.219\\
V & 765 & $-$0.211 & 0.002 & 4.28 & 0.06 & 0.229\\
J & 420 & $-$0.200 & 0.004 & 4.13 & 0.09 & 0.222\\
H & 399 & $-$0.201 & 0.004 & 3.98 & 0.10 & 0.225\\
K & 370 & $-$0.195 & 0.004 & 4.03 & 0.10 & 0.233\\
\hline
&&& SDSS $R_{23}$ Relation \\
B & 765 & $-$0.271 & 0.003 & 3.57 & 0.06 & 0.277\\
B$_o$ & 765 & $-$0.247 & 0.002 & 3.96 & 0.05 & 0.251\\
V & 765 & $-$0.259 & 0.002 & 3.64 & 0.05 & 0.262\\
J & 420 & $-$0.173 & 0.003 & 5.03 & 0.08 & 0.165\\
H & 399 & $-$0.172 & 0.003 & 4.95 & 0.08 & 0.166\\
K & 370 & $-$0.173 & 0.003 & 4.85 & 0.09 & 0.179\\
\enddata
\end{deluxetable}

\end{document}